%% file: paper.tex
\documentclass[conference,a4paper,10pt,times]{IEEEtran}
\IEEEoverridecommandlockouts
\usepackage{cite}
\usepackage{amsmath,amssymb,amsfonts}
\usepackage{algorithmic}
\usepackage{graphicx}
\usepackage{textcomp}
\usepackage{xcolor}
\def\BibTeX{{\rm B\kern-.05em{\sc i\kern-.025em b}\kern-.08em
    T\kern-.1667em\lower.7ex\hbox{E}\kern-.125emX}}

\input{custom}

\begin{document}

\title{\tool{} for DOPed Applications: A Compiler for Automated Data-Oriented Programming}

\author{
\IEEEauthorblockN{Jannik Pewny}
\IEEEauthorblockA{%
    \textit{Ruhr-Universit\"at Bochum, Germany}\\
jannik.pewny@rub.de}
\and
\IEEEauthorblockN{Philipp Koppe}
\IEEEauthorblockA{%
    \textit{Ruhr-Universit\"at Bochum, Germany}\\
philipp.koppe@rub.de}
\and
\IEEEauthorblockN{Thorsten Holz}
\IEEEauthorblockA{%
    \textit{Ruhr-Universit\"at Bochum, Germany}\\
thorsten.holz@rub.de}
}

\maketitle

\begin{abstract}
\input{sections/abstract}

\end{abstract}

\begin{IEEEkeywords}
Data-Oriented Programming, Exploitation, Compiler, Steroids, Slang
\end{IEEEkeywords}

\input{main}

\bibliographystyle{IEEEtran}
\bibliography{bibliography}

\input{sections/appendix}
\end{document}

%% file: custom.tex
\usepackage{graphicx}
\usepackage{enumerate}
\usepackage{amssymb}
\usepackage{url}
\usepackage{tabularx}
\usepackage{amsmath}
\usepackage{listings}

\usepackage{flushend}

\usepackage{caption}
\DeclareCaptionType{copyrightbox}
\usepackage{colortbl}
\usepackage{multicol}
\usepackage{xcolor}
\usepackage{paralist}

\usepackage{multirow}

\usepackage{subfig}

\usepackage{pifont} %
\newcommand{\xmark}{\ding{55}}

\usepackage{glossaries}
\setacronymstyle{long-short}
\loadglsentries{glossary}

\glsdisablehyper

\usepackage[inline]{enumitem}

\renewcommand{\subsubsection}[1]{{\parindent0pt\par{\textbf{#1.}}}}

\newcommand{\sidecomment}[1]{\marginpar{\scriptsize #1}}

\definecolor{ForestGreen}{RGB}{34, 138, 34}

\newcommand{\outline}[1]{\sidecomment{\textcolor{ForestGreen}{}}}

\lstnewenvironment{todoblock}
{\lstset{
	numbers = none,
   	numberblanklines=false,
   	frame=none,
   	caption={},
   	label={},
   	captionpos={},
	language={},
	tabsize=3,
	breaklines=true,
	basicstyle=\small\rmfamily\color{blue},
}\textcolor{blue}{TODO:}}
{}

\lstdefinelanguage{slang}
{
  morekeywords={
    func,
    call,
    if,
    else,
    while,
    loop,
    from,
    to,
    call,
    break,
    continue,
    return,
    for
  },
 keywordstyle=\color{blue}\ttfamily,
 stringstyle=\color{red}\ttfamily,
 commentstyle=\color{ForestGreen}\ttfamily,
 morecomment=[l][\color{ForestGreen}]{\#},
 morestring=[b]"
}

\makeatletter
\lst@AddToHook{TextStyle}{\let\lst@basicstyle\ttfamily\normalsize}
\makeatother

\definecolor{AquaMarine}{rgb}{0.5, 1.0, 0.83}
\definecolor{Mahogany}{rgb}{0.75, 0.25, 0.0}
\definecolor{CadetBlue}{rgb}{0.37, 0.62, 0.63}

\newcommand{\tool}{\textsc{Steroids}}
\newcommand{\ourlanguage}{\textsc{Slang}}

%% file: glossary.tex
\newglossaryentry{dop-asm}
 {
  name={DOP assembly}, 
  description={A list of DOP-Operations. Can be HL (may contains DOP-Operations not immediately available in the application) and LL (contains only DOP-Operations present in the application; DOP-Gadgets). Lowered Slang.}
 }

\newglossaryentry{dop-gadget}
 {
  name={DOP gadget}, 
  description={A DOP-Operation that is supported by the application.}
 }

\newglossaryentry{dop-op}
 {
  name={DOP operation}, 
  description={An assembly-like Instruction; e.g. mov, inc, add;}
 }

\newglossaryentry{hl-dop-asm}
 {
  name={High-Level DOP-Asm}, 
  description={DOP-Operations, which may not be immediately available in the application. Non-application-specific DOP-Assembly. Lowered Slang, can be further lowered to application-specifc DOP-Assembly.}
 }

\newglossaryentry{ll-dop-asm}
 {
  name={Low-Level DOP-Asm}, 
  description={DOP-Operations, which are available as DOP-Gadgets in an application. Lowered High-Level DOP-Asm.}
 }

\newglossaryentry{dop-script}
 {
  name={DOP script}, 
  description={The attacker's intent, expressed as a Slang- or DOP-Assembly-Script.}
 }

\newglossaryentry{dop-program}
 {
  name={DOP program}, 
  description={A data-structure, which triggers the DOP-Gadgets of an application to execute the DOP-Script}
 }
\newglossaryentry{dop-exploit}
 {
  name={DOP program}, 
  description={...}
 }

\newglossaryentry{gadget-definition}
 {
  name={gadget definition}, 
  description={Contents of an application's configuration. Defines, how to set a buffer's content to trigger a certain DOP-Gadget.}
 }

\newglossaryentry{recipe}
 {
  name={recipe}, 
  description={Entry in the Op-Graph; one way to synthesize another DOP-Op using other DOP-Ops.}
 }

\newacronym{acv}{ACV}{attacker-controlled value}
\newacronym{aslr}{ASLR}{address space layout randomization}
\newacronym{byose}{BYOSE}{bring your own scripting engine}
\newacronym{cfg}{CFG}{control-flow graph}
\newacronym{cfi}{CFI}{control-flow integrity}
\newacronym{cpi}{CPI}{code-pointer integrity}
\newacronym{coop}{COOP}{counterfeit-object oriented programming}
\newacronym{cots}{COTS}{commercial off-the-shelf}
\newacronym{dcr}{DCR}{destructive code-read}
\newacronym{dep}{DEP}{data-execution prevention}
\newacronym{dfi}{DFI}{data-flow integrity}
\newacronym{doa}{DOA}{data-oriented attack}
\newacronym{doe}{DOE}{data-only exploits}
\newacronym{dop}{DOP}{data-oriented programming}
\newacronym{jitrop}{JIT-ROP}{just-in-time return-oriented programming}
\newacronym{jop}{JOP}{jump-oriented programming}
\newacronym{op-graph}{Op-Graph}{Operations Graph}
\newacronym{ret2libc}{ret2libc}{return-to-libc}
\newacronym{rop}{ROP}{return-oriented programming}
\newacronym{rop-chain}{ROP chain}{return-oriented programming chain}
\newacronym{rop-gadget}{ROP gadget}{ROP gadget}
\newacronym{tlb}{TLB}{translation lookaside buffer}
\newacronym{wit}{WIT}{write-integrity testing}
\newacronym{xom}{XOM}{execute-only memory}

%% file: sections/abstract.tex
The wide-spread adoption of system defenses such as
the randomization of code, stack, and heap
raises the bar for code-reuse attacks.
Thus,
attackers utilize a scripting engine in target programs like a web browser
to prepare the code-reuse chain, e.g.,
relocate gadget addresses or perform a just-in-time gadget search.
However,
many types of programs do not provide such an execution context that an attacker can use.
Recent advances in \gls{dop} explored an orthogonal way to abuse memory corruption vulnerabilities and demonstrated that an attacker can achieve
Turing-complete computations without modifying code pointers in applications.
As of now, constructing \gls{dop} exploits requires a lot of
manual work---for every combination of application and payload anew.

In this paper, we present novel techniques to automate the process of generating
\gls{dop} exploits.
We implemented a compiler called \tool{}
that leverages these techniques and compiles our high-level language \ourlanguage{}
into low-level \gls{dop} data structures driving malicious computations at run time.
This enables an attacker to specify her intent in an application- and vulnerability-independent
manner to maximize reusability.
We demonstrate the effectiveness of our techniques and prototype implementation 
by specifying four programs of varying complexity in \ourlanguage{}
that calculate the Levenshtein distance,
traverse a pointer chain to steal a private key,
relocate a ROP chain, and perform a JIT-ROP attack.
\tool{} compiles each of those programs to low-level
\gls{dop} data structures targeted at five different
applications including GStreamer, Wireshark, and ProFTPd,
which have vastly different vulnerabilities and \gls{dop} instances.
Ultimately, this
shows that our compiler is versatile,
can be used for both 32-bit and 64-bit applications, works
across bug classes, and enables highly expressive attacks
without conventional code-injection or code-reuse techniques
in applications lacking a scripting engine.

%% file: main.tex
\input{sections/introduction}

\input{sections/background}

\input{sections/approach}
\input{sections/dop-scripts}

\input{sections/targets.tex}

\input{sections/evaluation}

\input{sections/discussion}

\input{sections/related-work}

\input{sections/conclusion}

\section*{Acknowledgment}
\input{sections/acknowledgements}

%% file: sections/introduction.tex
\section{Introduction}

\outline{1. Motivation}
\outline{1.1 Security Research focused on control-flow}
Attackers have to overcome more and more obstacles to exploit an application given
that defense mechanisms such as stack protections, \gls{dep}, \gls{aslr} and
\gls{cfi}~\cite{cfi,cfi-for-cots,ccfir} are nowadays widely deployed.
Modern defenses often require that an attacker adapts her exploit to the
current state of the application she is attacking.
For this reason, many modern exploits target browsers or PDF readers,
which come with a built-in scripting engine.
Utilizing the computational capabilities of these scripting engines,
e.\,g., JavaScript or Flash,
the attacker can repeatedly leverage a memory error, probe memory, find gadgets
just-in-time, and perform complex computations
to make the code-reuse chain compatible to the target application's security model.
However, system defenses like
\gls{xom} and its relatives~\cite{xnr,hidem,heisenbyte,near},
CPI~\cite{cpi} and Readactor~\cite{readactor,readactor-plus-plus}
aim to mitigate even advanced just-in-time code-reuse attacks.

\outline{1.2 Data-only attacks}
An alternative line of attacks targets non-control data~\cite{data-only}.
These so-called \glspl{doa} can have consequences
just as severe as code-reuse attacks, but neither alter code pointers
nor rely on the present code randomization. %
\outline{1.2.3 defenses impractical}
Despite \glspl{doa} being less well-explored than code-reuse attacks,
there have been efforts to ensure data integrity~\cite{wit, data-flow-integrity, valueguard, hard, data-randomization, data-space-randomization, salads, randomizing-data-structure}.
Nevertheless, efficient mitigation of \glspl{doa} with strong security guarantees
remains a challenging and open problem.

\outline{1.3 DOP}
\outline{1.3.1 What is DOP?}
Data-oriented programming (DOP) is an extension of \gls{doa},
which manipulates non-control data to use the application's own operations
to perform arbitrary computations of the attacker's choice.
Thus, they could provide the execution context modern code-reuse methods
need, even in applications which lack a scripting engine.
\outline{1.3.2 pressing topic}
We dub this a \emph{\gls{byose}} attack,
which is more versatile and potentially even more harmful
than \glspl{doa} alone,
and arguably makes securing data flow an even more pressing topic.

\outline{1.3.3 State-of-the-art}
The state of the art in \gls{dop}
is mostly concerned with searching for \glspl{dop-gadget}~\cite{dop}
or considers rather simple programs~\cite{block-oriented}.
However, using \glspl{dop-gadget} for non-trivial
\glspl{dop-exploit} is left to manual work,
as the resulting exploits are highly application- and vulnerability-dependent.
\outline{1.3.4 state-of-the-art is incomplete, not automated; instead, we...}
While existing work showed that diverse and powerful \glspl{dop-gadget} are available,
it is unclear which \glspl{dop-gadget} are actually
necessary or can be
expressed through others, or which minimal sets of \glspl{dop-gadget} achieve
useful expressiveness in different \gls{dop} attack modes.

In this paper, we close this gap: we provide novel insights into the flexibility of \gls{dop}, introduce a high
degree of automation for the construction of \gls{dop} exploits, and demonstrate
that complex attack payloads work across different applications and vulnerabilities.
More specifically, we present a high-level programming language called \ourlanguage{},
which can be used to describe \glspl{dop-program}.
\outline{2.2 Input: available DOP-gadgets}
Using descriptions of the \glspl{dop-gadget} available from a specific
vulnerability of a specific application
(conceptually, the output of the gadget search~\cite{automatic-data-oriented,dop}),
we use an automated way to craft the data structures that trigger \glspl{dop-gadget}
to carry out the computation described in the \gls{dop-program}.
Since the \ourlanguage{}~scripts are designed to be application-independent,
the available \glspl{dop-gadget} usually do not immediately yield all
necessary operations to express the attacker's intent.
To tackle this problem,
we support and encode \textit{\glspl{recipe}} for various operations,
i.\,e., ways to combine them
to express either one another or more high-level operations.
Implicitly, this creates a graph structure, which we call an \emph{\gls{op-graph}}.
Given such an \gls{op-graph}, we %
translate the attacker's script %
to use
only \glspl{dop-gadget}
present in
the target application.
As such, most of the manual effort to create a \gls{dop-exploit} for one
application can be reused to exploit other vulnerabilities
or applications with different sets of \glspl{dop-gadget}.

\outline{Teaser on Eval}
\outline{3.1 multiple apps, multiple scripts}
To demonstrate the practicality of our techniques, we implemented our approach in a tool called \tool{}. It can automatically build different
\glspl{dop-exploit} for different applications with different defense mechanisms.
E.\,g., one of our scripts expects to be run on a randomized binary~\cite{sok-automated-software-diversity},
and searches gadgets at runtime to ultimately mount a just-in-time code-reuse attack.
This effectively bypasses the protection offered by, say, Binary Stirring~\cite{binary-stirring} or Compiler-assisted Code Randomization~\cite{ccr}.
One key element is that the attacker's script does not have to be modified to
be compiled for different vulnerabilities or different applications.
\outline{3.2 assemble new ops, branch-free, interactive}
Furthermore,
our experiments include examples to highlight various features of our compiler,
namely
(i) to \textit{bootstrap} additional \glspl{dop-gadget},
(ii) to compile to \textit{branch-free} code to compensate for lacking conditional jumps, and
(iii) to support an \textit{interactive mode} if only a single \gls{dop-gadget} can be executed at a time.

\outline{4. Result}
\outline{4.1 DOP is not esoteric, but reliably exploitable}
Ultimately, our results show that \gls{dop} is not esoteric,
but can indeed be utilized to reliably execute complex attack payloads in target programs.
\outline{4.2 hard/impossible to exploit apps are back on the menu}
These BYOSE attacks transfer the possibility of advanced code-reuse attacks, such as JIT-ROP, to target applications
without a built-in scripting engine.

\smallskip \noindent
In summary, our main contributions are as follows:

\begin{itemize}
\outline{5.3 Lowered DOP-Assumption}
    \item We provide novel insights regarding the minimal requirements for successful \gls{dop} exploitation. We can push the boundaries by bootstrapping new \gls{dop} instructions, utilizing branch-free code and an interactive mode.
\outline{5.1 Automation / Reuse}
    \item We develop novel techniques to automate the construction of
\glspl{dop-exploit} and enable the reuse of exploits across target applications and vulnerabilities.
\outline{5.2 Prototype}
    \item We present our prototype implementation~\tool{}, which can compile exploits specified in our high-level scripting language \ourlanguage{}~for a given target application.%
\outline{5.4 DOP: see 4.1}
    \item We demonstrate the practicality of our techniques and prototype by implementing complex programs in \ourlanguage{}~and automatically compiling them to \glspl{dop-program} in binary form. Our \ourlanguage{}~scripts include ROP chain relocation and a runtime gadget search. We show that BYOSE enables just-in-time code-reuse attacks for applications \emph{without} a built-in scripting engine and thus, extends the set of potential targets.
\end{itemize}

%% file: sections/background.tex
\section{Technical Background}

Before diving into the details of our approach, we briefly introduce
the necessary technical background information on data-oriented programming needed to understand the
building blocks of our method.

\subsubsection{Code-reuse attacks}
The motivation for return-to-libc attacks~\cite{ret2libc},
\gls{rop}~\cite{rop} and its variants~\cite{jop,rop-journal,blind-rop,rop-without-returns}
was the wide-spread adoption of the $W \oplus X$ policy,
which prohibits the execution of writable data sections
rendering code injection infeasible. %
These so called code-reuse attacks have in common that they reuse pieces of existing code,
which are called ROP gadgets,
and employ a mechanism to chain these gadgets by creating or modifying code pointers. %
Code-randomization defenses~\cite{heisenbyte-15,heisenbyte-18,binary-stirring,aslr-linux,timely-rerandomization,xifer,tanenbaum-rerandomization,function-shuffling} shuffle and modify gadgets, forcing attackers to relocate their ROP chain
or even search new gadgets to construct a new ROP chain on the fly (JIT-ROP~\cite{jit-rop,info-leak}).
To mitigate just-in-time code-reuse attacks,
many defenses have been proposed that hide code and code pointers~\cite{readactor,
readactor-plus-plus,near,isomeron,shuffler,oxymoron} or aim to preserve the control flow~\cite{cfi,cfi-for-cots,
cpi,ccfir,aslr-guard}.

\subsubsection{Data-only attacks}
Leveraging memory corruptions to modify non-control data is an orthogonal attack vector~\cite{data-only,automatic-data-oriented,new-doa}.
The attacker corrupts data structures or data pointers
to modify the data flow.
In this way, the program's logic can be tricked into leaking sensitive information like private
keys~\cite{automatic-data-oriented}
or passing attacker-controlled
content to access control data structures or critical functions such as \texttt{execve}.
The existing efforts to prevent \glspl{doa}~\cite{wit, data-flow-integrity, valueguard, hard, data-randomization, data-space-randomization, salads, randomizing-data-structure}
impose a high performance overhead or do not provide strong security guarantees.

\subsubsection{Data-oriented programming}
Similar to ROP, \Gls{dop} aims to achieve a high degree of expressiveness,
but without modifying code pointers.
The attacker utilizes a memory corruption vulnerability to modify existing data structures
or to inject new data structures.
The ultimate goal of \gls{dop} is to repurpose the logic of existing pieces
of code (\gls{dop} gadgets) to perform the attacker desired computation (\gls{dop} operations).
However, arbitrarily chaining of gadgets spread across the program,
as is possible with ROP, is not possible for \gls{dop}.
A certain proximity in the control-flow graph is required to ensure
continuous execution and selection of the next \gls{dop} gadget.
Conceptually,
two methods to chain \gls{dop} gadgets exist:
First, the \textit{non-interactive mode} utilizes an existing
loop in the program with \gls{dop} gadgets in its loop body to ensure continuous execution.
Additionally,
a mechanism is necessary that selects the next data structure like a
linked list or an array of structs.
Often the loop condition needs to be modified to keep the loop running,
which can be accomplished with the memory corruption or the execution of the first \gls{dop} gadget.
Since the payload contains the trigger for the memory corruption and the data for all instructions,
the attacker needs no further interaction with the target application.
Second, in the \textit{interactive mode},
the attacker initiates the loop iterations separately by
repeatedly leveraging the memory corruption
and feeding the data structures that trigger the desired \glspl{dop-gadget}.
The motivation for this mode is that the attacker may not be able to
extract state of the target program, e.g.,
if the memory corruption has no read capabilities.
In this case adapting a code-reuse payload on the attacker side is severely
hampered.
That being said, if the attacker can extract state,
she can \textit{adapt} following \glspl{dop-gadget} for an even more
efficient \textit{adaptive interactive mode}.
For the remainder of the paper, we will assume that this is not the case, though.\\

\noindent
We define the following three steps to launch a DOP attack:
\begin{description}
\item[Gadget search:] Find \gls{dop} gadgets that are reachable directly or indirectly by the memory corruption vulnerability.
The scope depends on the bug,
but can range from structures on the active stack frame to the whole program.
Furthermore,
the outputs of \gls{dop} gadgets must be readable by the following \gls{dop} gadgets.
This step also involves searching gadget dispatcher loops or establish an outside mechanism to chain gadgets.
\item[DOP instance setup:] Collect path constraints to reach the basic blocks containing \gls{dop} gadgets and
setup data structure templates that trigger desired \gls{dop} gadgets.
In the non-interactive mode, one also needs to setup the mechanism that feeds the required data structures and triggers the next cycle.
\item[Payload preparation:] The native gadgets of the \gls{dop}
instance often provide constrained and unusual operations.
For example, some \gls{dop} instances lack a controllable program counter,
conditional operations, or even basic arithmetic and data movement.
Thus,
this step involves bootstrapping a convenient set of operations
and creating a stream of operations that compensates lacking features.
Finally,
the collection of data structure contents and inputs for the memory corruption must be compiled.
\end{description}

Previous work~\cite{dop} focused on the automation of the gadget search.
While they also performed DOP instance setup and payload preparation,
they did so in a mostly manual fashion.
In contrast, Block-Oriented Programming~\cite{block-oriented}
provides a higher degree of automation,
but its program synthesis is NP-hard.
Thus, it can mainly solve the constraints for \glspl{dop-program} that are not too complex,
and rather belongs to the step of DOP instance setup.

In this paper, we pursue orthogonal research
by focusing on automating the payload preparation.
We leave the gadget search to related work and
interface the DOP instance setup.
That is, we define a file format for this stage and process it accordingly
in the payload preparation.
The key design idea to avoid NP-hardness is to work
with unconstrained data flow and inputs in later steps,
i.\,e., the gadgets are arbitrarily stitchable and work for all inputs.
Almost all manual work that is left to the attacker belongs
to her custom payload or
to the DOP instance setup, which can be seen as retargeting our prototype
compiler~\tool{} to a new ``platform''.

%% file: sections/approach.tex
\section{Automated DOP Exploit Compiler}

In this section,
we describe the techniques that enable automated compilation of \glspl{dop-program}
under consideration of constrained \glspl{dop-gadget}, %
and provide a high-level overview of~\tool{}'s
components and their interactions.

\subsection{Assumptions and Attacker Model}

Our assumptions are aligned with previous research and we adopt the
general scenario given in the previously reported \gls{dop} instances:
First,
we assume that the target application is protected with DEP,
ASLR, and advanced control-flow hijack mitigations such as fine-grained CFI or
fine-grained load-time code randomization. %
\outline{Stack-Overflow}
Second, we assume the presence of a memory corruption vulnerability.
Our test cases make use of both stack-based vulnerabilities
and heap-based vulnerabilities (see Section~\ref{targets}).
\outline{Triggering input}
Lastly, we assume that an input triggering the vulnerability is known.

\subsection{Overview}
\begin{figure}[tb]
\centering
  \includegraphics[width=0.48\textwidth]{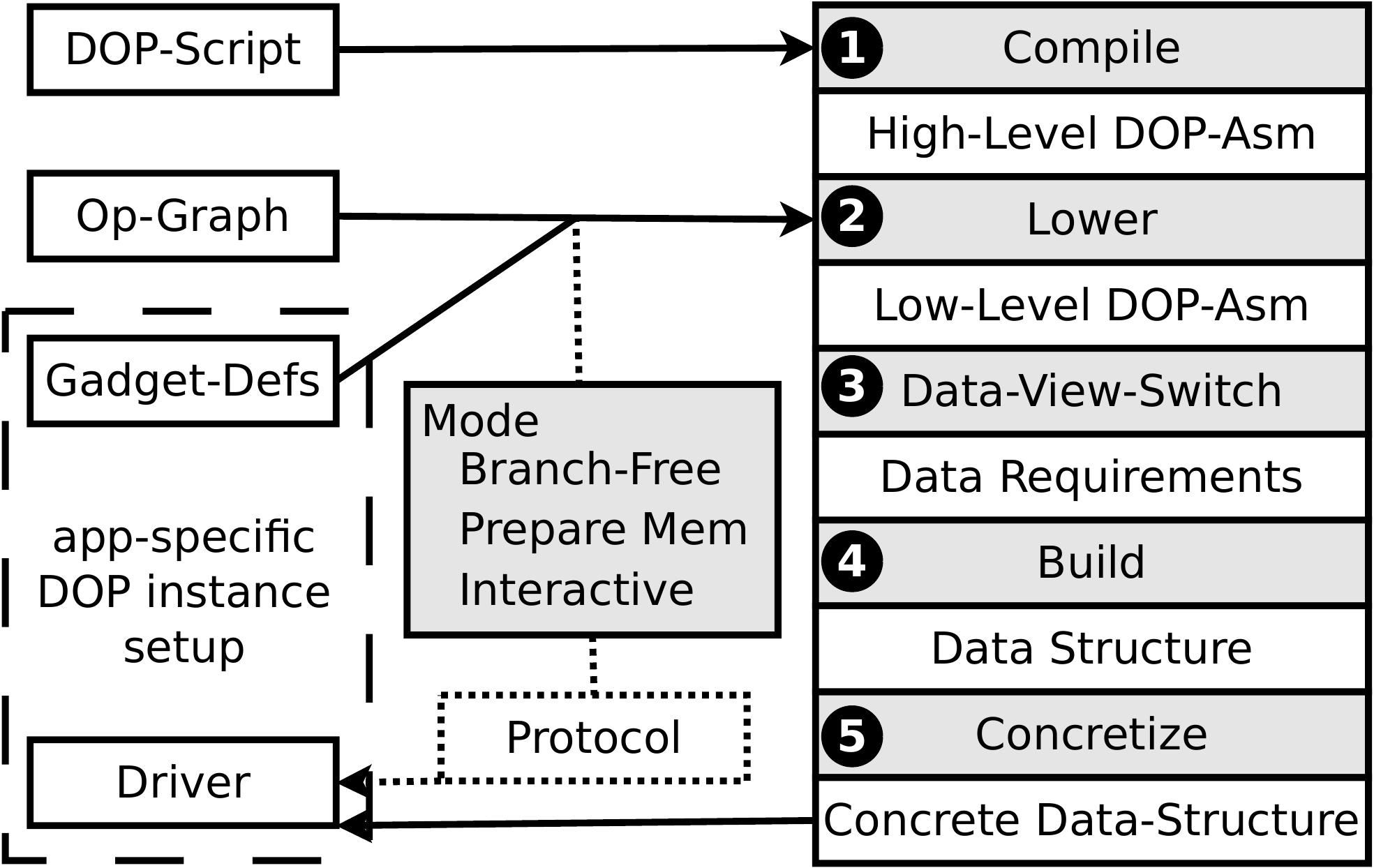}
  \caption{Workflow of \tool{}}
  \label{figure-workflow}
\end{figure}

Figure~\ref{figure-workflow} provides an overview of
the inputs, stages, and intermediary artifacts of our approach.
Note that we focus on the concepts first
and give examples for the artifacts and more detail about their format
later in this section.

The dashed rectangle on the bottom left
represents the DOP instance setup with its two components:
The \textit{\glspl{gadget-definition}} and the \textit{driver}.
The \glspl{gadget-definition} specify the necessary constraints for the
data structures driving the execution of a specific \gls{dop-gadget}
present in the application.
Naturally, these constraints are highly specific to both the application and
the vulnerability.
Conceptually, they directly result from findings of the gadget
search~\cite{automatic-data-oriented,dop,block-oriented},
but are set manually in our approach.
However, to correctly judge the necessary effort,
one has to keep in mind that a \gls{gadget-definition} in the five target
applications we evaluate (see Section~\ref{targets}) is on average only about ten lines long,
and that one only needs to define about half a dozen of them (see Section~\ref{eval-implemented-gadgets}).
In context of compilers, the \glspl{gadget-definition} are analogous to the
instructions of a new architecture.

The \textit{driver} represents the interface to the vulnerable application (e.\,g., it writes to a file or opens a TCP connection
to trigger the initial vulnerability).
Given that this
is highly specific to the exploit and
the application's input stream, %
we again rely on the attacker to supply the \textit{driver}.

Now follows the centerpiece of this paper, the payload preparation.
\outline{Slang to DOP-Asm}
We separate the automatic compilation of \gls{dop} exploits into five stages,
which are reflected in the design of our prototype implementation
(see Figure~\ref{figure-workflow}). %
After the attacker specified her intent using our high-level programming language,
\ourlanguage{},
we first \textit{compile}~(\ding{182}) her \gls{dop-script} into an assembly-like
format: \emph{\gls{hl-dop-asm}}.
This \gls{hl-dop-asm} is independent of the application,
that is, it may use \glspl{dop-gadget} which the application does not provide.

\outline{Gadgets plus Op-Graph leads to LL-DOP}
In step~\ding{183}, %
we use a set of \glspl{recipe},
which encode how to express \glspl{dop-gadget} through other \glspl{dop-gadget}.
We use these recipes and the \gls{op-graph} they implicitly define
through their interdependencies,
to \textit{lower} the \gls{hl-dop-asm} into \gls{ll-dop-asm}.
The latter uses only \glspl{dop-gadget} from the \glspl{gadget-definition},
i.\,e., only \glspl{dop-gadget} available for this specific vulnerability and application.
For the most part, the \glspl{recipe} and \gls{op-graph}
can be reused for multiple applications,
unlike the application-specific \glspl{gadget-definition}.

\outline{LL-DOP-Asm to Data Requirements}
The \gls{ll-dop-asm} is meant to be read with execution in mind,
but we have to think %
in terms of data driving the execution
in context of \gls{dop}. %
Thus, in the \textit{data-view-switch}~(\ding{184}), %
we use the content of the \glspl{gadget-definition}.
This results in the \textit{data requirements},
which are the constraints for a data structure
that would execute the \gls{dop-script}
if it were placed in the vulnerable application.

\outline{Data Requirements to Data Structure}
In step~\ding{185},
we actually \textit{build} the \textit{data structure} fulfilling
the formerly created \textit{data requirements}.
\outline{Concretization}
\outline{(Altered constants, relocated addresses)}
Since this involves potentially expensive constraint-solving,
the \textit{data structures} can contain placeholders instead of concrete values.
An additional \textit{concretization}~(\ding{186}) step
can then quickly replace constants or addresses leaked at runtime.

\outline{Driver; conceptually, not done by us}
Lastly, we pass the final data structure back to the \textit{driver},
in order to execute it in the target application.

\subsection{Modes}

The \textit{mode} element in Figure~\ref{figure-workflow} represents an accumulation of compiler flags triggering different program transformations, and
indicates the \gls{dop} gadget supply mode required for the specific target \gls{dop} instance.
There are three separate, but partially interlocking optional modes:
branch-free, memory preparation, and interactive.

\subsubsection{Branch-Free}
\stepcounter{lstlisting}
If one cannot synthesize conditional jumps for the application,
one cannot easily express %
if/else statements.
Listing~\arabic{lstlisting} shows how
to use arithmetics\footnote{This example uses an equality operator, which produces a one if the values are equal and a zero otherwise. Alternatively, one can
place alternative values in an array and compute different indices,
or use conditional \glspl{dop-gadget}.}
to express case distinction~\cite{Z3-turing-complete}:
All operations are executed on the right side,
whereas one needs to conditionally skip operations on the left side.
That is, the right side can do without a \lstinline{conditional goto}.
This transformation of an operation must fulfill two properties:
First, it must only have an effect, if the \lstinline{here}-bit is set and
second, it may only assume that the values it uses are properly initialized,
if the \lstinline{here}-bit is set.

Thus, we %
compute a \lstinline{here}-bit at the start of every basic blocks
by comparing a \lstinline{state}-variable to a basic block's ID.
Then, we can simply change the \lstinline{state}-variable instead of using \lstinline{goto}s.
The result is that
the execution sequence of basic blocks is no longer important,
only that they are executed often enough.\\

Listing~\arabic{lstlisting}: Conditional and Branch-Free Code\\
\begin{minipage}{.15\textwidth}
\lstset{language={},
        basicstyle=\footnotesize,
        keywordstyle=\color{blue}\ttfamily,
        stringstyle=\color{red}\ttfamily,
        commentstyle=\color{ForestGreen}\ttfamily,
        morecomment=[l][\color{magenta}]{\#},
        numberblanklines=false,
        frame=lrtb,
        label={listing-conditional},
        captionpos=t,
        tabsize=4,
        mathescape
}
\begin{lstlisting}
t = x + y
if(z == 3):
	r = t
\end{lstlisting}
\end{minipage}\hfill
\begin{minipage}{.27\textwidth}
\lstset{language={},
        basicstyle=\footnotesize,
        keywordstyle=\color{blue}\ttfamily,
        stringstyle=\color{red}\ttfamily,
        commentstyle=\color{ForestGreen}\ttfamily,
        morecomment=[l][\color{magenta}]{\#},
        numberblanklines=false,
        frame=lrtb,
        label={listing-branch-free},
        captionpos=t,
        tabsize=4,
        mathescape
}
\begin{lstlisting}
t = x + y
here = z == 3
r = here * t + (1-here) * r
\end{lstlisting}
\end{minipage}

\subsubsection{Memory Preparation}
When \glspl{dop-gadget} are placed in separate buffers,
e.g., in the interactive mode,
it may not be possible to place persistent variables next to the
\glspl{dop-gadget} using them.
However, one can use \glspl{dop-gadget} to place the variables somewhere
in memory, before the remainder of the \gls{dop-script} is executed.
This phase therefore collects all variables, translates their initial values
into immediate values used by \glspl{dop-gadget},
and transforms the script in order
to use the addresses the variables are written to.

\subsubsection{Interactive}
A \lstinline{goto} \gls{dop-gadget} is not necessary
in the interactive mode, because the attacker can decide
which operation to send next.
However, this also places the burden to decide which operation to send next
on the attacker.
This is inherently problematic if the execution state is not known to the
attacker, e.\,g., if her \gls{dop-script} requires case distinctions or loops.
Luckily, due to the branch-free transformation,
it only matters that an operation is executed often enough,
because additional executions have no effect.
Thus, one big loop would theoretically suffice to execute any program,
e.\,g., if the \glspl{dop-gadget} are saved in a linked list,
which the attacker can corrupt to form a circle.
However,
executing only what is needed is obviously more performant.

For reducible programs~\cite{reducible},
we can automatically generate a \textit{protocol}
which details how often to execute which sequence of basic blocks.
Using well-known techniques, we analyze the CFG,
inline functions, and dissect the program into the building blocks
of structured programming: sequences, which are executed in-order,
selections (if/else), for which both the true path and the
false path are executed once, and iterations, for which
we rely on the attacker for annotations in the \gls{dop-script}
to hint the number of repetitions.

\subsection{Artifacts}
Now that we have discussed the stages of our approach,
let us give more details on the used artifacts.

\subsubsection{High-Level Language: \ourlanguage{}}
Our tool \tool{} provides a high-level language for the development of
\glspl{dop-program}.
Our \tool{} programming language, \ourlanguage{}, provides
\begin{itemize}
  \item compound expressions
  \item typed variables (addresses, bytes, int16, int32...)
  \item type-sensitive arrays
    \begin{itemize}
      \item constant initialization: strings, int-arrays, hex-dumps
      \item automatic length-variable for constant initialization
    \end{itemize}
  \item structured control-flow
    \begin{itemize}
      \item if/else blocks
      \item loops: for, while, repeat $\hdots$ until, infinite
      \item break/continue
    \end{itemize}
  \item (recursion-free) procedures
\end{itemize}

This is in contrast to Qool~\cite{Q-exploit-hardening} for \gls{rop}
or MinDOP~\cite{dop} for \gls{dop},
which provide only ``list-of-statements'' languages.
Since the exemplary scripts in Section~\ref{dop-scripts}
show that our language is rather straightforward and conventional,
we refrain from showing the formal grammar of \ourlanguage{} in this paper,
both for brevity and to omit technicalities.

The compilation from \ourlanguage{} to \gls{hl-dop-asm} itself is also
straightforward (e.\,g., the elements of structures control flow are compiled
to checks, labels and \lstinline{goto}s).
Compound expressions are parsed and compiled into single two-operand
operations for each node in the parse tree,
using temporary variables, if necessary.
Array accesses multiply the index with the size of an array element,
before using the result as an offset after dereferencing the array's address.

\subsubsection{\gls{dop-asm}}
While \ourlanguage{} is used to define the attacker's intent,
\gls{dop-asm} is used as an intermediary step,
but also as
a convenient language for the \glspl{recipe} %
and for inline-use in \ourlanguage{}.
It is
low-level and assembly-like,
where each line holds a command with its operands.
Additionally, it features
\begin{itemize}
  \item typed variables (addresses, bytes, int16, int32...)
  \item conditional operations
  \item labels as jump-targets
  \item macros, e.\,g.,
    \begin{itemize}
      \item to generate unique labels or variable names
      \item to reserve only one temporary variable, if a gadget is used multiple times
      \item to perform compile-time computation
    \end{itemize}
  \item compiler directives, e.\,g., to tell the compiler
    \begin{itemize}
      \item to advance a program counter
      \item to increase the size of a packet
    \end{itemize}
\end{itemize}

For a simple example, refer to Listing~\ref{listing-add-recipe}.
Note that the computation
could also be done byte-wise, such that one requires only $2^8 * 4$ instead of $2^{32}$ loop iterations. 

\lstset{language={},
        basicstyle=\footnotesize,
        keywordstyle=\color{blue}\ttfamily,
        stringstyle=\color{red}\ttfamily,
        commentstyle=\color{ForestGreen}\ttfamily,
        morecomment=[l][\color{magenta}]{\#},
        numberblanklines=false,
        frame=lrtb,
        caption={\Gls{recipe} to synthesize an \lstinline{add}, using \lstinline{mov, dec, inc} and a \lstinline{conditional goto}. Suffixes for the bit-width of the operations and operands are omitted.},
        label={listing-add-recipe},
        captionpos=t,
        tabsize=4,
        mathescape,
}
\begin{lstlisting}
add dst src
	int cpy
	mov cpy src
	:start
		if_zero_goto cpy :end
		dec cpy
		inc dst
	goto :start
	:end
\end{lstlisting}

\subsubsection{\Acrlong{op-graph}}
\Glspl{recipe} encode how to synthesize higher-level operations from simpler ones.
The \gls{op-graph} is created simply by parsing
a set of \glspl{recipe} and drawing edge sets from an operation to the operations
used in a specific recipe.

\begin{figure}[tb]
\centering
  \includegraphics[width=0.38\textwidth]{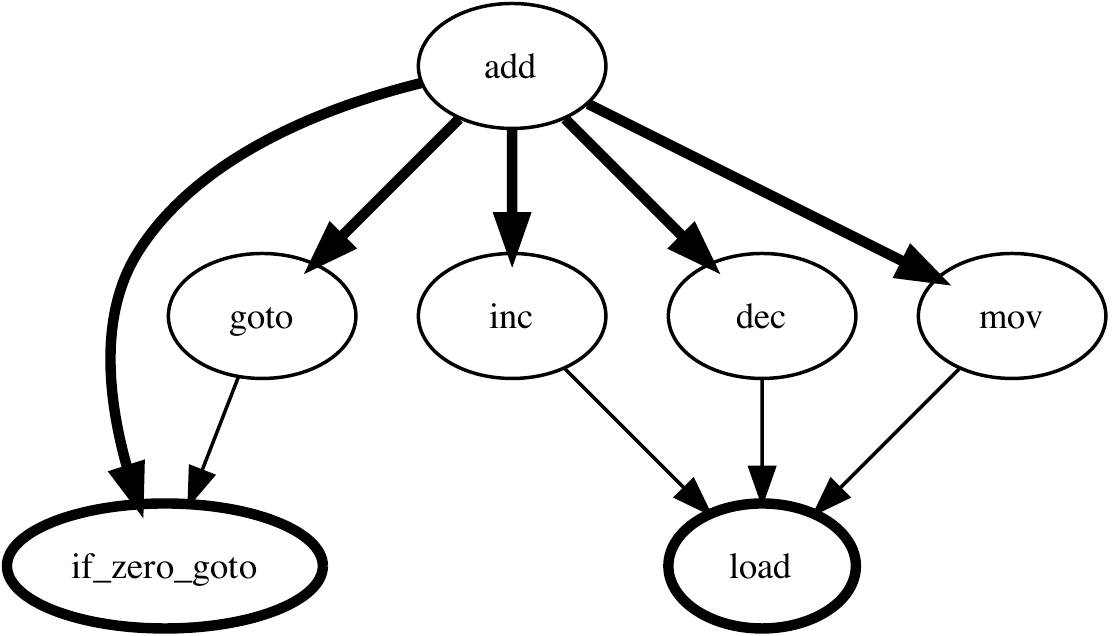}
  \caption{\gls{op-graph} corresponding to the \gls{recipe} from Listing~\ref{listing-add-recipe} for synthesizing an \lstinline{add}.
}
  \label{figure-op-graph}
\end{figure}

To illustrate our approach, %
a \gls{recipe} to express the \lstinline{add} \gls{dop-op} (\lstinline{*p += *q})
is given in Listing~\ref{listing-add-recipe}.
The \gls{op-graph} in Figure~\ref{figure-op-graph} includes the dependencies
defined in the recipe, and shows that
\lstinline{add} uses five other \glspl{dop-op},
indicated by the bold edges.
Furthermore, it shows
that the application only has two \glspl{dop-gadget}:
\lstinline{load} (\lstinline{*p = **q}) and a \lstinline{conditional goto},
indicated by the bold nodes.
A recursive graph search on the \gls{op-graph} starting at the \lstinline{add}-node,
shows each \gls{dop-op} in this recipe requires only other \glspl{dop-op},
which themselves require only \glspl{dop-gadget} present in the application.
Thus, this recipe can provide an
\lstinline{add} \gls{dop-op} to the attacker,
even though the application does not have an \lstinline{add} \gls{dop-gadget}.

A \lstinline{goto} can be synthesized using a
\lstinline{conditional goto} with an always-true dummy condition,
and a \lstinline{mov} (\lstinline{*p = *q}) can be implemented using a \lstinline{load}
with one additional indirection,
but expressing \lstinline{inc/dec} through a \lstinline{load} may be a little
surprising:
In Section~\ref{eval-wireshark} we describe how to use a lookup table
to achieve this feature.

The lowering of \gls{hl-dop-asm} to \gls{ll-dop-asm}
requires only macro expansion along the found paths to the \glspl{dop-gadget}:
We successively substitute a \gls{dop-op} with the body of its recipe,
taking care to substitute \gls{recipe}'s parameters with the \gls{dop-op}'s arguments,
until we end up using only \glspl{dop-gadget}.
These are then expanded using their \glspl{gadget-definition}
to form the data requirements.

For convenience,
if a \gls{dop-op} cannot be synthesized because of missing \glspl{dop-gadget},
our compiler generates an And/Or-graph from the recipes,
which reports all possible alternatives
of which \glspl{dop-op} one would have to implement to complete said recipes.
Furthermore,
we designed our \glspl{op-graph} to be combinable trivially,
so simple file concatenation allows
reusing \glspl{recipe} for other applications.

\subsubsection{DOP-Gadget Definition}
In \gls{dop}, one uses skillfully crafted data structures to drive the
program to perform specific operations.
Thus, at one point, one has to switch from the execution-perspective
to the data structure-perspective that is actually used in the application
(see Listing~\ref{listing-simple-dop-gadget} for an example).
The language we use to define data structures is fairly simple:
It supports symbols and constants, which may be initialized to specific
addresses and values, memory dereferences and offsets,
where the latter offers a typed array-index formulation as syntactic sugar.

\lstset{language={C},
        basicstyle=\footnotesize,
        keywordstyle=\color{blue}\ttfamily,
        stringstyle=\color{red}\ttfamily,
        commentstyle=\color{ForestGreen}\ttfamily,
        numberblanklines=false,
        frame=lrtb,
        caption={Artificial example of a \gls{dop-gadget}.},
        label={listing-simple-dop-gadget},
        captionpos=t,
        tabsize=4,
        mathescape
}
\begin{lstlisting}
struct s { int *value, *from, *to; };
struct s *ptr = ...;
// memory corruption of ptr
if(ptr->value == 3)
	*(ptr->to) =  *(ptr->from);
\end{lstlisting}

\begin{figure}[h]
\centering
  \includegraphics[width=0.30\textwidth]{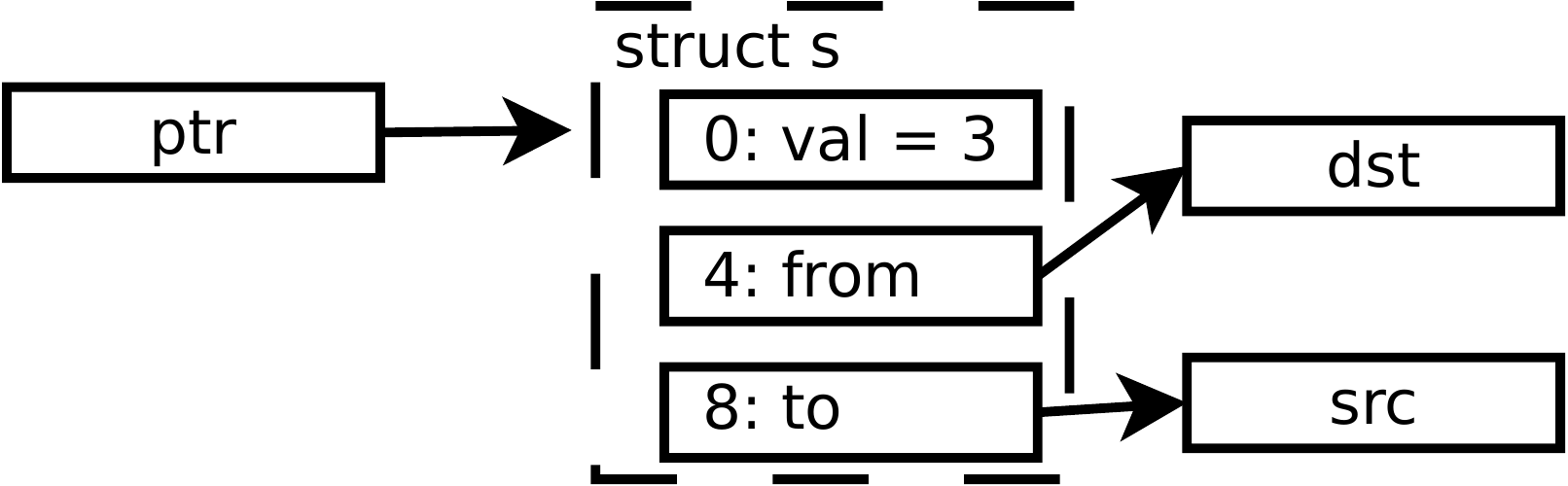}
  \caption{Data structure to invoke Listing~\ref{listing-simple-dop-gadget}'s \gls{dop-gadget}.}
  \label{figure-data-structure}
\end{figure}

\input{scripts/data-structure-definition.tex}

For example, Listing~\ref{listing-data-structure-definition} defines the
data structure from Figure~\ref{figure-data-structure} to trigger a
\lstinline{mov} operation.
These data requirements are then transformed into constraints for the Z3 SMT Solver:
We require that, at a certain position in the buffer,
there is an address (\lstinline{ptr}).
Somewhere else in the buffer, there must be the value \lstinline{3},
and \lstinline{ptr} must point to it.
This value must be followed by another pointer (\lstinline{from}),
which must have the value given by the attacker as the second parameter (\lstinline{src}),
and yet another pointer (\lstinline{to}),
which must have the value of the first parameter (\lstinline{dst}).

While it is not necessary for this simple example,
our compiler can handle multiple dereferences and offsets on both sides
of the equation,
e.\,g., to write to a member in a struct.

\subsubsection{Protocol}
The protocol holds nested sequences of basic blocks and, if they occur in a loop,
optionally how often to execute that loop.
We opted for S-Expressions as an easy-to-parse format.
E.\,g., we translate the protocol
\vspace{-1.5mm}
\begin{center}
$(BB_1, (3, BB_2, BB_3), BB_4)$ 
\end{center}
\vspace{-3mm}
into the trace
\vspace{-1.5mm}
\begin{center}
$BB_1, BB_2, BB_3, BB_2, BB_3, BB_2, BB_3, BB_4$.
\end{center}

Note that a protocol is only required in the interactive mode,
where \glspl{dop-gadget} can be combined arbitrarily.
Thus, the protocol is not part of the \textit{data requirements}
and only needs to be passed to the \textit{driver}.

\subsection{Implementation Details}
We implemented~\tool{} on Linux in Python and use the Z3 SMT Solver
to solve the constraints of the \textit{data requirements}.
We compiled \glspl{dop-program} for 32-bit and 64-bit Linux target applications.
Conceptually, it should also work for other operating systems since the
generated data structures by themselves are not OS-dependent.

\subsubsection{Optimizations}
~\tool{} is meant to provide maximum flexibility in its compile targets,
since essentially every \gls{dop}-instance offers different \glspl{dop-gadget},
and we therefore chose generality over efficiency.
Still, when emitting \gls{dop-asm},
~\tool{} tries to use the simplest \gls{dop-op}.
E.\,g., by
emitting \lstinline{inc/dec} instead of \lstinline{add/sub},
or by
avoiding the oftentimes rather expensive
\lstinline{neq/eq} \glspl{dop-gadget} when comparing against a constant,
in favor of calculating
the condition of a conditional \gls{dop-gadget} at compile time.

Furthermore, we allow annotations for weighted edges in the \gls{op-graph}.
By default, a \gls{dop-gadget} native to the target application has a weight
of one and a \lstinline{goto} a weight of ten to account for repeated execution
of a loop-body.
Our compiler can then sums the weights of all the used \glspl{dop-gadget}
and pick the \glspl{recipe} with the lowest cumulative weight.
~\tool{} also reduces swapping when lowering complex expression,
and it does not add or subtract zero.
Furthermore, it tries to avoid multiplying by one,
e.\,g., when computing the indices to an array access,
if the array is defined as a byte array.

%% file: scripts/data-structure-definition.tex
\lstset{language={},
        basicstyle=\footnotesize,
        keywordstyle=\color{blue}\ttfamily,
        stringstyle=\color{red}\ttfamily,
        commentstyle=\color{ForestGreen}\ttfamily,
        morecomment=[l][\color{magenta}]{\#},
        numberblanklines=false,
        frame=lrtb,
        caption={\Gls{gadget-definition} / Data structure requirements for the data structure from Figure~\ref{figure-data-structure} to feed the \gls{dop-gadget} from Listing~\ref{listing-simple-dop-gadget}. The address of \lstinline{ptr} is an input to the compiler, while \lstinline{src/dst} are the parameters of the \gls{dop-gadget}.},
        label={listing-data-structure-definition},
        captionpos=t,
        tabsize=4,
        mathescape
}
\begin{lstlisting}
mov dst src
	ptr deref offset(0) = 3
	ptr deref offset(4) = src
	ptr deref offset(8) = dst
\end{lstlisting}

%% file: sections/dop-scripts.tex
\section{DOP Scripts}
\label{dop-scripts}

This section presents a selection of high-level exploit payloads implemented in \ourlanguage{}.
They not only demonstrate the capabilities of \gls{dop} and~\tool{}
regarding expressiveness and complexity of payloads,
but also security implications, because they feature %
ROP chain relocation and on-the-fly gadget search
against targets \emph{without} a built-in scripting engine.

\input{scripts/levenshtein.tex}

\subsection{A classic algorithm: Levenshtein distance}
\outline{description levenshtein}
\outline{why chosen}
We chose
the Levenshtein distance~\cite{yujian2007normalized} as an introductory example,
because it is a well-known algorithm and allows to assess the
compilation and execution times for the applications.
It computes the edit distance between two strings, i.\,e.,
the number of operations necessary to transform one string into the other.
\outline{implementation}
In particular, our \ourlanguage{} implementation in Listing~\ref{listing-levenshtein}
features:
\begin{enumerate*}
  \item Nested for-loops, requiring checking values for equality and conditional jumping.
  \item 2D-array operations, i.\,e., memory access uses indices, which are computed using multiplication.
  \item Numeric comparison and conditional execution to compute the minimum of three numbers.
  \item Functions to compute minimum and indices.
\end{enumerate*}

\subsection{SSL pointer-chain dereference}
\outline{description ssl deref chain}
In OpenSSL, the data structure for saving an SSL private key is fairly complicated:
From a base object (\texttt{ssl\_ctx}),
one has to successively follow a chain of eight pointers at different offsets
in their respective data structures,
to finally reach the secret key (see Listing~\ref{listing-ssl-deref-chain}).
\outline{why chosen}
We adapted this example from Hu et. al~\cite{dop},
because it shows the use of a high-level description for a non-trivial
\gls{byose}-attack. %
Naturally,
not every application uses SSL private keys and such a base object
may not always be withing immediate reach in every application.
However,
this is still a realistic attack, which is impossible without either an
adaptive multi-step attack leaking data or a scripting engine.
Note that one should not judge the script's complexity by it's brevity:
Our experiments will show that executing it without optimization can require
thousands of \glspl{dop-gadget}.

\input{scripts/ssl-deref-chain.tex}

\subsection{ROP-chain Relocation}
\outline{description relocator}
To account for ASLR, one can relocate a \gls{rop-chain},
which means adjusting the addresses of the \glspl{rop-gadget} in the
exploit buffer using a dynamically retrieved address.
Attackers usually utilize a built-in scripting engine to perform this step,
but this example accomplishes this task using \gls{dop},
demonstrating that this kind of attack can be applied to target
applications without a built-in scripting engine.
As Listing~\ref{listing-relocator} shows,
the attacker merely provides the address of a code pointer
(\lstinline{addr_of_code_pointer}),
like a return address on the stack or a function pointer,
and the offset between the code pointer and the image base
(\lstinline{offset_to_base}).
Furthermore, she prepares
the exploit buffer using the offset between the image base and the
\glspl{rop-gadget}, as it would be in a non-randomized address layout.
Again, the brevity indicates rather \ourlanguage{}'s expressiveness
than the script's complexity:
It may very well require executing thousands of \glspl{dop-gadget}.

\newpage
\input{scripts/relocator.tex}

\lstset{language={},
        basicstyle=\footnotesize,
        keywordstyle=\color{blue}\ttfamily,
        stringstyle=\color{red}\ttfamily,
        commentstyle=\color{ForestGreen}\ttfamily,
        morecomment=[l][\color{magenta}]{\#},
        numberblanklines=false,
        frame=lrtb,
        caption={An exemplary ROP chain to execute \lstinline{execve("/bin/sh")} in 32-bit Linux.},
        label={listing-rop-chain-buffer},
        captionpos=t,
        tabsize=4,
        mathescape
}
\begin{lstlisting}
0x00: //bi					0x18: &buffer
0x04: n/sh					0x1C: &(pop ecx; ret)
0x08: 00 00 00 00			0x20: &buffer + 8
0x0C: &(pop eax; ret)		0x24: &(pop edx; ret)
0x10: 0B 00 00 00			0x28: &buffer + 8
0x14: &(pop ebx; ret)		0x2C: &(int 0x80)
\end{lstlisting}

\subsection{\Gls{jitrop}}
The relocation approach from the previous example does not work anymore,
if a binary is protected with fine-grained code
randomization such as \textit{Binary Stirring}~\cite{binary-stirring},
because most of the previously known ROP gadgets are eliminated.
Instead,
an attacker can employ \gls{jitrop} to dynamically scan the process memory to search required gadgets on-the-fly,
and assemble the ROP chain in the exploit buffer accordingly.
Our implementation probes the program's code memory for specific gadgets
and then uses their locations in a freshly generated ROP chain.
Just like the other examples,
the Slang script in Listing~\ref{listing-jitrop} %
can be compiled and executed for interactive and non-interactive mode \gls{dop} instances without changes.

\input{scripts/jit-rop.tex}

%% file: scripts/levenshtein.tex
\lstset{language={slang},
        basicstyle=\footnotesize,
        numberblanklines=false,
        frame=lrtb,
        title={~},
        caption={Slang-code to compute the Levenshtein distance.},
        label={listing-levenshtein},
        captionpos=t,
        tabsize=4,
        mathescape
}
\begin{lstlisting}
byte[]  s = "kitten"
byte[]  t = "sitting"

byte[56] d

func idx(int the_row, int the_col, int width)
	idx = the_row * width + the_col

func min(int a, int b, int c)
	min = a
	if(b < min)
		min = b
	if(c < min)
		min = c

for j from 0 to s.length + 1
	call idx(0, j, s.length + 1)
	d[idx] = j

for i from 0 to t.length
	call idx(i + 1, 0, s.length + 1)
	d[idx] = i + 1

	for j from 0 to s.length
		int j_plus_1 = j + 1

		call idx(i, j + 1, s.length + 1)
		int del_cost = d[idx] + 1

		call idx(i + 1, j, s.length + 1)
		int ins_cost = d[idx] + 1

		call idx(i, j, s.length + 1)
		int sub_cost = d[idx]
		if(s[j] != t[i])
			sub_cost += 1

		call min(del_cost, ins_cost, sub_const)
		call idx(i + 1, j + 1, s.length + 1)
		d[idx] = min
\end{lstlisting}

%% file: scripts/ssl-deref-chain.tex
\lstset{language={slang},
        basicstyle=\footnotesize,
        numberblanklines=false,
        frame=lrtb,
        caption={Slang-code to retrieve a private key from SSL.},
        label={listing-ssl-deref-chain},
        captionpos=t,
        tabsize=4,
        mathescape
}
\begin{lstlisting}
# The address of (or offset to) ssl_ctx is
# given as a parameter to the compiler.
addr p -> ssl_ctx

int[] offsets = [10 0 4 20 24 0 0]

for i from 0 to offsets.length
	p =* p
	p += offsets[i]

# At this point, p points to the private key and can
# be copied to an address of the attacker's choice.
\end{lstlisting}

%% file: scripts/relocator.tex
\lstset{language={slang},
        basicstyle=\footnotesize,
        numberblanklines=false,
        frame=lrtb,
        caption={Slang code to relocate a ROP chain.},
        label={listing-relocator},
        captionpos=t,
        tabsize=4,
        mathescape
}
\begin{lstlisting}
byte[] rop_chain = {...}

# A ROP-chain holds data, mixed with addresses
# of ROP-gadgets. This array holds the offsets
# of those addresses in the ROP-chain buffer.
int gadget_offsets[] = [0x0C 0x14 0x24 0x28 0x2C]

addr image_base =* addr_of_code_pointer
image_base -= offset_to_base

# Relocate each gadget.
for i from 0 to gadget_offsets.length
	rop_chain[gadget_offsets[i]] += image_base
\end{lstlisting}

%% file: scripts/jit-rop.tex
\lstset{language={slang},
        basicstyle=\footnotesize,
        numberblanklines=false,
        frame=lrtb,
        caption={Slang-code to scan the memory for \glspl{rop-gadget}.},
        label={listing-jitrop},
        captionpos=t,
        tabsize=4,
        mathescape
}
\begin{lstlisting}
# pop eax, pop ebx, pop ecx, pop edx, int 0x80
byte[] g_bytes   = {58 c3 5b c3 59 c3 5a c3 cd 80}
int[]  g_lengths = [2 2 2 2 2]
int[]  g_offsets = [0x0C 0x14 0x24 0x28 0x2C]

addr start_addr =* addr_of_code_pointer

func is_gadget(addr a, byte[] gadget, int g_length)
	for i from 0 to g_length
		if(gadget[i] != a[i])
			return 0
	return 1

byte[] cur_g_bytes = g_bytes
for j from 0 to g_offsets.length
	addr at = start_addr
	loop
		call is_gadget(at, cur_g_bytes, g_lengths[j])
		if(is_gadget == 1)
			rop_chain[g_offsets[j]] = at
			break
		at += 1
	cur_g_bytes += g_lengths[j]
\end{lstlisting}

%% file: sections/targets.tex
\section{Applications with DOP Instances}
\label{targets}

We compiled and executed the example \glspl{dop-script} above for the
five different applications we present in this section.
Note that we only crafted the first of them:
The other four were not ``homemade'' by us.
Table~\ref{table-comparison} provides a comparison of the \gls{dop} instances
with respect to the available \glspl{dop-gadget}.

\input{tables/comparison.tex}

\subsection{Interpreter}
\outline{intro}

\outline{its actually dop}
This bytecode interpreter is an exemplary application,
which loads a file filled with bytecode into memory and then interprets the embedded instructions.
Thus,
this ~\gls{dop} instance is compatible with the non-interactive gadget chaining mode,
although it does not require the exploitation of a memory error.

\outline{dop-gadgets}
The interpreter has three arithmetic instructions:
addition, subtraction, and multiplication.
It also has an instruction to move data,
and a \lstinline{conditional goto} to modify the control-flow.
Since the application internally uses an instruction-counter,
both a \lstinline{goto} and a \lstinline{calculated goto} can be implemented by modifying its value.

The interpreter's comparably rich set of \glspl{dop-gadget}
is by far not complete:
E.\,g., there are no comparison operators for (in)equality.
Also, since the operand size is fixed to 32 bit, it is not immediately suitable to work on single bytes.
Most importantly, this \gls{dop} instance lacks \lstinline{load/store} gadgets.

\outline{synthesis}
{\parindent0pt\par{\textbf{Challenges.}}}
To interact with memory outside the \gls{dop} instance,
we bootstrapped the \lstinline{load/store} gadgets using \textit{self-modifying \gls{dop}}.
Since the data structures for all \glspl{dop-gadget} are already in memory,
an operation can alter the operands of other operations.
We leverage a \lstinline{mov} to overwrite the source or destination of another
\lstinline{mov} with values computed at runtime.
Thus, we can simulate native \lstinline{load/store} gadgets.
Note that \lstinline{mov} is the ``weakest'' of the three basic data movement
\glspl{dop-op}, and thus,
any of the three can be used to synthesize the other two.
This however,
may not hold without self-modifying \gls{dop}.
As we will see in the Section~\ref{evaluation},
this application shows that \glspl{dop-gadget} such as \lstinline{add} and \lstinline{mul}
are very important for the efficiency of the overall \gls{dop}-instance.

\subsection{Wireshark}
\label{eval-wireshark}
The packet analyzer Wireshark
suffered from a stack-buffer overflow
(CVE-2014-2299), which ultimately results in packet contents overflowing
local variables (specifically \lstinline{cinfo}) and parameters
(specifically \lstinline{packet_list}).
Listing~\ref{listing-wireshark} shows the relevant lines for this application's
\glspl{dop-gadget}.

Detailed descriptions are available in the literature~\cite{dop,hardscope},
so we only give a brief summary of triggering the \glspl{dop-gadget} here.
Setting specific values to struct members of \lstinline{packet_list} and \lstinline{cinfo} enables
\lstinline{mov/load/store} and \lstinline{conditional inc} \glspl{dop-gadget}.
However, the condition and the target of the \lstinline{conditional inc}
are not independent of one another, as they stem from the same value.
Thus,
one has to set up a fake \lstinline{packet_list} data structure in memory,
and use a \lstinline{mov/inc/mov}-sequence to increment an arbitrary memory address.
Additionally,
a fake linked-list pointing at itself has to be employed to create an endless loop
and keep the \gls{dop} instance running.
Furthermore,
the file position indicator serves as a virtual instruction pointer:
manipulating its value results in a non-linear sequence of packets being read and enables a \lstinline{goto}.
The exploit payload is given at one go,
but the data structures are being loaded to memory one by one.
Thus, this \gls{dop} instance has properties of interactive and non-interactive modes:
single-shot exploit, native \lstinline{goto}, but no self-modifying \gls{dop}.

\lstset{language={C},
         basicstyle=\footnotesize,
         keywordstyle=\color{blue}\ttfamily,
         stringstyle=\color{red}\ttfamily,
         commentstyle=\color{ForestGreen}\ttfamily,
         numberblanklines=false,
         frame=lrtb,
         caption={Partial Wireshark source code showing the lines for mov/load/store and conditional inc \glspl{dop-gadget}.},
         label={listing-wireshark},
         captionpos=t,
         tabsize=4,
         mathescape
 }
\begin{lstlisting}
record = packet_list->physical_rows[row];
record->col_text[col] =
         (gchar *) cinfo->col_data[col];
if(!record->col_text_len[col])
   	++packet_list->const_strings;
\end{lstlisting}
\lstset{language={}}

{\parindent0pt\par{\textbf{Challenges.}}}
Arithmetic operations are severely hampered since
the only arithmetic \gls{dop-gadget} is an \lstinline{inc},
but expressive programs can still be compiled and the resulting
\glspl{dop-program} are efficient enough to pose a security threat.

To synthesize an \lstinline{add} using the recipe given in Listing~\ref{listing-add-recipe},
we implemented an 8-bit \lstinline{dec} using a 256-byte lookup
table. %
This table is placed at a 256-byte aligned address,
and its $i$th value is initialized to $i-1 \text{ mod } 256$.
Now, we can write the value we want to decrement into the lowest byte of
the address to the lookup table, and load the value saved there.
The implication of this construct is severe: We crafted an arithmetic \gls{dop-op}
using only data-movement \glspl{dop-op}.

\subsection{Gstreamer}
\label{eval-gstreamer}
The Gstreamer multimedia framework suffered from a heap-corruption
vulnerability (CESA-2016-0004)
in its decoder for FLIC files, which are GIF-like animations.
In particular, the vulnerability repeatedly allows to write arbitrary
byte-sequences to arbitrary, positive offsets.
The decoding happens in a fresh thread of a fresh process and since a thread's heap
is usually aligned to 64\,MB, the last three bytes of the addresses of
multiple relevant heap objects are basically static.
Evans published an exploit for this vulnerability and the write-up contains
lots of technical details~\cite{gstreamer-exploit}.
The author makes heavy use of partial pointer overwrites
(namely, the last three bytes to keep the unknown first bytes intact)
to modify the addresses used by two \lstinline{memcpy}s to move data arbitrarily,
to dereference pointers and even to abuse the frame-time calculation to perform
addition.
To do so, a fake \lstinline{GstPad}-instance is created to avoid leaving the main loop,
and data is copied into the input buffer to create new \glspl{dop-gadget},
i.\,e., self-modifying \gls{dop} is used.
Ultimately, a code-reuse attack is launched by crafting a call to the
\lstinline{system}-function.
Listing~\ref{listing-gstreamer} shows the details relevant for \gls{dop}.

Unlike the other applications,
this one does not use a 32-bit address space,
and thus shows that \tool{} can also deal with 64-bit
applications.
Furthermore, this \gls{dop} instance does not rely on a stack corruption, but
leverages a heap corruption, demonstrating that \tool{} works across
bug classes.

{\parindent0pt\par{\textbf{Challenges.}}}
While the exploit by Evans is certainly nifty, we have to go significantly
beyond the published write-up.

In GStreamer, data is first read in 4\,KB-sized chunks into a linked-list of
\lstinline{GstMemory}-objects.
A class called \lstinline{GstAdapter} is then responsible to merge these
chunks of memory to accommodate the size the decoder actually needs.
In this specific decoder, every frame is freshly allocated and therefore,
their addresses become more unpredictable over time.
Evans did not have to deal with this,
because the entire exploit fit into the first 4\,KB-chunk.
However, reallocation is not acceptable for larger \glspl{dop-program}:
Our examples easily grew to 90\,KB.
Thus, we bootstrap our \glspl{dop-program}
by first copying their main content into the frame-buffer and then
creating a fake \lstinline{GstAdapter}-object to trick
the application into thinking that there is only one big memory object.
This not only facilitates reliable frame-wise self-modifying \gls{dop},
but also allows to modify the \lstinline{skip-, size-} and \lstinline{assembled_len}-variables
of the \lstinline{GstAdapter}-object to redirect \gls{dop} execution to arbitrary frames.
Conceptually, this still belongs to the steps of gadget search
and \gls{dop} instance setup to facilitate the execution of \glspl{dop-gadget},
which is not in this paper's scope.
We thus do not automate this task.
In practice however, we used our flexible compilation and \gls{recipe}
mechanism to make it considerably easier.
The second challenge is the addressing.
The instructions that place constant data use relative offsets, but the \lstinline{memcpy}s
naturally require absolute addresses. Furthermore, the predictable addresses are
all in the three byte-vicinity around the frame-buffer.
Thus, a second bootstrap-step dynamically discloses the first
bytes of the heap-objects and fixes the addresses of all indirectly addressed
\gls{dop}-variables.
Note that our compiler generates both these steps automatically.

As for the actual \glspl{dop-gadget}, the
\lstinline{mov-, load-} and \lstinline{add-}instructions were already described
by Evans~\cite{gstreamer-exploit} and are rather straightforward.
However, the arbitrary \lstinline{store}-gadget was more challenging, because
the native code does not contain the required data flow. We constructed
the \lstinline{store}-gadget by overwriting operands of a \lstinline{mov}-gadget
at run time utilizing self-modifying \gls{dop}. While \lstinline{conditional goto}
is not a strict requirement for Turing-complete computation, it helps to avoid
unreasonable large payload sizes for complex \gls{dop} programs. We use the fact
that the FLIC-decoder skips frames with the wrong header ID and we leverage the 
fake \lstinline{GstAdapter} and self-modifying \gls{dop} to skip frames by
modifying their header IDs in order to construct a \lstinline{conditional goto}-gadget.

\input{./scripts/gstreamer}

\lstset{language={}}

\subsection{Mini-Server}
Hu et al.~\cite{dop} modeled this application after an FTP-Server to serve
as an example for \gls{dop}. However, please note that it is in no way tailored
to our compiler.
The application suffers from a stack-based buffer
overflow when reading data from a socket into a local buffer.
Invoking the three \glspl{dop-gadget} given in Listing~\ref{listing-mini-server}
is fairly straightforward.
However, one has to take care to preserve local variables,
i.\,e., to reset \lstinline{connect_limit} with each \gls{dop}-packet.
Also, the parameter \lstinline{buf} for \lstinline{readData} is pushed onto
the stack only once and needs to be restored each iteration.

{\parindent0pt\par{\textbf{Challenges.}}}
The \lstinline{mov} \gls{dop-gadget} cannot move all values, e.g.,
the value \lstinline{STREAM} triggers another gadget.
Thus,
we opted to use the \lstinline{load} to implement a general purpose
\lstinline{mov} for this application.
We also had to use a temporary scratch space for the \lstinline{conditional load},
because its \lstinline{else}-case clobbers some adjacent values.

This \gls{dop} instance lacks the very important \lstinline{store} gadget,
which is necessary to write to arrays and other memory locations.
Due to the interactive mode we cannot leverage self-modifying
\gls{dop} to synthesize \lstinline{store} from \lstinline{mov/load}.
Luckily,
the \lstinline{mov} and \lstinline{add}
\gls{dop-gadget} are executed in sequence in a single packet.
Thus, we can use the \lstinline{mov} to modify the parameters
of the \lstinline{add} \gls{dop-gadget}.
This yields a store-add combination: \lstinline{*(*p+4) += *q},
where the $4$ stems from the different offsets
for \lstinline{typ} and \lstinline{total}.
Adjusting that offset leads to a \lstinline{**p += *q} \gls{dop-gadget}.
To convert this into a pure \lstinline{store}, we first
use this \gls{dop-gadget} to add the value at \lstinline{**p} to itself,
effectively multiplying it by two.
In base two, this introduces a zero at the lowest bit.
By repeating this 32 times, we can store a zero at a location of our choice.
Finally, we invoke this \gls{dop-gadget} a 33rd time, adding *q to the zero,
which results in the desired \lstinline{store}-\gls{dop-gadget} (\lstinline{**p = *q}).
Again, the implication is severe: We used an arithmetic gadget (\lstinline{add})
to achieve data-movement\footnote{
Other constructs are possible, too. E.\,g., using only \lstinline{dec} and \lstinline{mul}:
to move $q$ to $p$, one multiplies $p$ by $0$, decrements it, multiplies it by $-1$ and finally multiplies it by $q$.}.

\lstset{language={C},
        basicstyle=\footnotesize,
        keywordstyle=\color{blue}\ttfamily,
        stringstyle=\color{red}\ttfamily,
        commentstyle=\color{ForestGreen}\ttfamily,
        numberblanklines=false,
        frame=lrtb,
        caption={Partial Mini-Server source code showing the \lstinline{conditional load}, \lstinline{mov}, and \lstinline{add} \glspl{dop-gadget}. \lstinline{size}, \lstinline{srv}, \lstinline{type} and \lstinline{connect_limit} are local variables.},
        label={listing-mini-server},
        captionpos=t,
        tabsize=4,
        mathescape
}
\begin{lstlisting}
while(connect_limit--) {
  readData(sockfd, buf);    // stack buffer overflow
  if(*type == NONE) break;
  if(*type == STREAM)       // conditional load
      *size = *(srv->cur_max);
  else {
      srv->typ = *type;     // mov
      srv->total += *size;  // add
}}
\end{lstlisting}

\subsection{ProFTPd}
An integer overflow, which results in a stack-based buffer overflow,
in the FTP server ProFTPd (CVE-2006-5815) allows the
attacker to write almost arbitrarily into the process' memory.
As a result, multiple \glspl{dop-gadget} are available,
which allow Turing-complete computation.
There are six functions involved and the example is well-documented
in the literature~\cite{hardscope,dop}, so we omit the vulnerable code.
ProFTPd allows an interactive \gls{dop}-mode,
but needs two packets per \gls{dop-gadget}.

{\parindent0pt\par{\textbf{Challenges.}}}
Triggering the
\lstinline{conditional mov} gadget was straightforward.
The \lstinline{add} \gls{dop-gadget} operates on a fixed global data structure,
so one has to first move the two operands into said data structure,
invoke the \lstinline{add}, and then fetch the result.
The \lstinline{load} and \lstinline{store} gadgets need to perform
the double-dereferencing in two steps,
which can be performed as a regular two-packet \gls{dop-gadget},
but one needs additional \lstinline{mov}s to place the parameters
accordingly.

\subsection{nginx}
A flaw in the chunked size parser enables the attacker to inject a very large size number.
Due to an unsafe unsigned to signed conversion that number can be interpreted as negative bypassing a buffer size check,
which leads to a stack-based buffer overflow (CVE-2013-2028).
Previous work~\cite{dop} analyzed this bug,
but ruled it out for \gls{dop}.
We achieved a proof-of-concept \gls{dop} instance, albeit with limited capabilities.

{\parindent0pt\par{\textbf{Challenges.}}}
The target data structure
\lstinline{ngx_http_request_t} is used in many places, is therefore
heavily constrained and clobbers many struct fields.
We resolved this by rarely moving the struct in memory and fulfilling the constraints with memory preparation.
Even though a \lstinline{conditional load} and \lstinline{byte inc}
are available, the interactive mode
impedes synthesizing \lstinline{mov/store}.
While this \gls{dop} instance can compute arithmetic operations and move values
among the limited number of write slots,
the \lstinline{store}-constraint prevented us from running complex \gls{dop} scripts.

%% file: tables/comparison.tex
\begin{table}
  \centering
  \caption{Comparison of the evaluated DOP-instances.\\
1: Implicitly, because the conditional command is present.\\
2: There is only one packet in memory at a time, thus one packet cannot modify the contents of the next packet.\\
3: Implicitly, through self-modifying \gls{dop}.\\
4: Synthesized from \lstinline{mov} and \lstinline{add} \gls{dop-gadget} sequence.\\
5: Implicitly, through interactive mode/branch-free code.
}
  \begin{tabular}{l|c|c|c|c|c}
                        & \rotatebox{70}{Interpreter} & \rotatebox{70}{Wireshark}       & \rotatebox{70}{GStreamer}    & \rotatebox{70}{Mini-Server}    & \rotatebox{70}{ProFTPd} \\
\hline
\textbf{Arithmetic} & & & & \\
inc (*p += 1)           &             \xmark          &       \phantom{1}$\checkmark^1$ &  \xmark                      &              \xmark            & \xmark \\
conditional inc & & & & \\
\multicolumn{1}{r|}{(if(...) *p += 1)}
                        &             \xmark          &                   \checkmark    &  \xmark                      &              \xmark            & \xmark \\
add (*p += *q)          &             \checkmark      &                   \xmark        &  \checkmark                  &              \checkmark        & \checkmark \\
sub (*p -= *q)          &             \checkmark      &                   \xmark        &  \xmark                      &              \xmark            & \xmark \\
mul (*p \texttt{*}= *q) &             \checkmark      &                   \xmark        &  \xmark                      &              \xmark            & \xmark \\
gte (*p = *p $\ge$ *q)
                        &             \checkmark      &                   \xmark        &  \xmark                      &              \xmark            & \xmark \\
\hline
\textbf{Movement} & & & & \\
mov (*p = *q)           &             \checkmark      &                   \checkmark    &  \checkmark                  &              \checkmark        & \phantom{1}$\checkmark^1$ \\
conditional mov & & & & \\
\multicolumn{1}{r|}{(if(...) *p = *q)}
                        &             \xmark          &                   \xmark        &  \xmark                      &              \xmark            & \checkmark \\
load (*p = **q)         &             \xmark          &                   \checkmark    &  \checkmark                  &  \phantom{1}$\checkmark^1$     & \checkmark \\
conditional load & & & &  \\
\multicolumn{1}{r|}{(if(...) *p = **q)}
                        &             \xmark          &                   \xmark        &  \xmark                      &              \checkmark        & \xmark     \\
store (**p = *q)        &             \xmark          &                   \checkmark    &  \phantom{3}$(\checkmark)^3$ & \phantom{24}$\checkmark^{2,4}$ & \checkmark \\
\hline
\textbf{Control-Flow} & & & & \\
goto                    & \phantom{1}$\checkmark^1$   &                   \checkmark    &  \checkmark                  & \phantom{5}$(\checkmark)^5$    & \phantom{5}$(\checkmark)^5$ \\
conditional goto        &             \checkmark      &                   \checkmark    &  \phantom{3}$(\checkmark)^3$ & \phantom{5}$(\checkmark)^5$    & \phantom{5}$(\checkmark)^5$ \\
calculated goto         &             \checkmark      &                   \checkmark    &  \xmark                      & \phantom{5}$(\checkmark)^5$    & \phantom{5}$(\checkmark)^5$ \\
\hline
\textbf{Mode} & & & & \\
interactive             &             \xmark          &                   \xmark        &  \xmark                      &              \checkmark        & \checkmark \\
self-modifying          &             \checkmark      & \phantom{2}$\text{\xmark}^2$    &  \checkmark                  & \phantom{24}$\checkmark^{2,4}$ & \xmark \\
  \end{tabular}
  \label{table-comparison}
\end{table}

%% file: scripts/gstreamer.tex
\lstset{language={C},
        basicstyle=\footnotesize,
        keywordstyle=\color{blue}\ttfamily,
        stringstyle=\color{red}\ttfamily,
        commentstyle=\color{ForestGreen}\ttfamily,
        numberblanklines=false,
        frame=lrtb,
        caption={Partial GStreamer source code showing reading from an arbitrary address to the frame buffer's start (\lstinline{dest}), placing of constant data at a chosen location in or after the frame buffer, writing to an arbitrary address, and the addition.},
        label={listing-gstreamer},
        captionpos=t,
        tabsize=4,
        mathescape
}
\begin{lstlisting}
flx_decode_delta_fli(...) {
  ...
  // read; attacker controls content of flxdec
  memcpy(dest, flxdec->delta_data, flxdec->size);
 
  // attacker also controls contents of data...
  start_line = (data[0] + (data[1] << 8));
  lines =  (data[2] + (data[3] << 8));
  data += 4;
  dest += (flxdec->hdr.width * start_line);

 while(lines--) {
   ...
   count = *data++;
   while(count --) {
     ...
     // ... which determines offset and written value
     *dest++ = *data++;
  }
 }
}
...
gst_flxdec_chain(...) {
  ...
  while(...) {
    // calls flx_decode_delta_fli
    flx_decode_chunks(...);
    ...
    // write; attacker controls content of flxdec
    memcpy(flxdec->delta_data,
           flxdec->frame_data, flxdec->size);
    ...
	// add
    flxdec->next_time += flxdec_frame_time;
  }
  ...
}
\end{lstlisting}

%% file: sections/evaluation.tex
\section{Evaluation}
\label{evaluation}

\outline{4 scripts, 5 apps}
To evaluate the capabilities of \tool{},
we have compiled each of our four showcase \glspl{dop-script}
against all five different applications from the last section.
\outline{table with numbers}
We then measured both compilation time and runtime,
the size of the generated \gls{dop-program}, and the number of
executed \glspl{dop-gadget}.
Influences on runtime,
results for optimized \glspl{dop-script},
and the number of necessary \glspl{dop-gadget} are also discussed.

\input{tables/table-non-interactive}
\input{tables/table-interactive}

\subsection{Quantitative Analysis}
We conducted our experiments on Ubuntu 18.04.1 LTS, using a single core of
an Intel i7 @ 2.9\,GHz with 4\,GB RAM.

For the Levenshtein algorithm, we measured the runtime of comparing a 7-character
string against an 8-character string.
Since not every application comes with SSL, we artificially added the 
relevant data structure into the program,
and measured how long it took to dereference it.
For the Relocator,
we measured how long it took to relocate a given \gls{rop-chain}.
In our JIT-ROP example, however, the runtime of the algorithm depends heavily
on where in memory the ROP gadgets are located.
For comparable results, we thus normalized the algorithms to
always scan exactly one kilobyte of memory.

Our results are summarized in Table~\ref{table-quantitative-non-interactive}
and Table~\ref{table-quantitative-interactive}.
\outline{compilation; linear, parallel}
The compile time scales roughly linear with the number of
\glspl{dop-gadget}.
One \gls{dop-gadget} takes about 150ms,
which is mostly spent by the SMT solver to create the data structures.
Naturally, more complex data structures have a tendency to take longer,
especially when they are nested or when there are arrays involved.
However, unless the available space for a data structure is very close to
the minimal space necessary, we found the compile time to be fairly stable.

Note that this step can oftentimes be parallelized,
if one decouples the compilation of single \glspl{dop-gadget}.
This
can also be used in the interactive
case, e.\,g., for GStreamer.
However, since we evaluated on a single core,
the additional overhead of starting separate processes may outweigh
the benefit of simpler constraint solving.
Furthermore, we noted that a noteworthy portion of a program's
\glspl{dop-gadget} are usually identical,
which makes caching an attractive time saver during development.

\outline{non-interactive: quick}
A comparison of the number of executed \glspl{dop-gadget}
in Table~\ref{table-quantitative-non-interactive}
(e.\,g., for the Levenshtein algorithm)
shows that the non-interactive mode enables efficient execution of \glspl{dop-program},
even if the underlying application offers only minimal arithmetic capabilities.
\outline{process-startup}
For the Interpreter, the runtime is even clearly dominated by the time to start
the process (e.\,g., for the Relocator).
\outline{interactive: network}
However, a single \gls{dop-gadget} in the interactive mode is roughly
three orders of magnitude slower,
because network-interaction dominates the runtime,
despite using a local loopback interface in our experiments.

The \gls{dop} instance in Wireshark is non-interactive,
but it reads the data structure for every single \gls{dop-gadget} from a file.
Hence, it spends a large amount of its runtime with file I/O.
Similarly, GStreamer invokes costly \lstinline{malloc}- and
\lstinline{memcpy}-operations for every \gls{dop-gadget},
even before the actual gadget is executed.
Nevertheless,
the biggest impact of runtime stems from the lack of powerful \glspl{dop-gadget}
for arithmetic, in particular for equality comparisons and multiplications,
which are utilized for array accesses.
E.\,g. a subtraction operation would remedy this situation.
This effect is most pronounced in the Wireshark example,
because it does not even have an efficient \lstinline{add},
which then has to be expressed through repeated \lstinline{inc}.

While the execution time naturally has a certain variance,
the chosen \glspl{dop-exploit} are deterministic
and there are many executed \glspl{dop-gadget}.
This seems to average out the
execution time per \gls{dop-gadget} in individual target applications,
especially since the runtime is usually dominated by factors other than computation.

These experiments show that \tool{} facilitates the automatic compilation
of \glspl{dop-exploit} from a high-level language.
Especially since the intermediary \gls{hl-dop-asm} and \gls{ll-dop-asm} scripts
are kept around, we argue that the generated programs can serve as an
even more useful starting point for hand-crafted optimization.

If one reduces the number of the aforementioned problematic operations in \ourlanguage{},
e.\,g, by using different variables instead of arrays
or using pointer-arithmetic instead of array-indices to reduce multiplications,
one can speed up the \gls{dop-program} considerably
(see last column of Table~\ref{table-quantitative-interactive}).
Especially when the number of loop iterations is known beforehand,
the costly equality-check against a variable can be avoided.
Instead one compares with a constant,
which can often be implemented with a much cheaper conditional \gls{dop-op} (see last column of Table~\ref{table-quantitative-non-interactive}).
For a small and fixed number of iterations one can also unroll the loop to avoid
the comparison altogether.

In the interactive gadget chaining mode, the attacker may not know the current state, e.g.,
specific values of variables,
of the \gls{dop} instance during runtime.
Consequently,
computed branches are not supported,
because the \textit{Driver} on the attacker machine has no means
to evaluate branch decisions and select the next \gls{dop-gadget}.
This branch-free transformation of the code has a huge performance impact.
First, it increases the number of \glspl{dop-op}.
Second, loops cannot be exited early, because the attacker does not know
whether the loop-condition is met and must therefore iterate,
until an upper approximation of loop iterations is reached.
This is especially critical for nested loops (e.\,g., in Levenshtein or JIT-ROP).

\subsection{Implemented \Glspl{dop-op}}
\label{eval-implemented-gadgets}
\Glspl{dop-op} fall into three separate categories:
First, the \glspl{dop-gadget}, which are implemented as \glspl{gadget-definition},
require the data-view of \gls{dop}.
They are a little tricky to implement and they are highly specific to both the
application and the vulnerability, so that it is unlikely that they can be
reused. Luckily, one has to implement only very few of them.
As Table~\ref{table-nof-gadgets} shows, a new target application requires
about five of these with an average length of 8.6 lines of code.
GStreamer is the clear exception, because only the \gls{dop-op} to write
constant data is implemented in this way.

Second, application-specific \glspl{dop-op},
which are implemented in \gls{ll-dop-asm}.
They also have to deal with the oddities of the application,
e.\,g, clobbered adjacent values, input values for which they do not work or
application-specific variables.
It is important, however, that these \glspl{dop-op} do not export such behavior,
i.\,e., they define clean \gls{hl-dop-asm}.
Since the execution-view is more familiar,
and since one deals with expected, albeit cumbersome behavior,
they are much easier to implement.
However, it is still unlikely that such a recipe can be reused
for another application.
There are again only about five of these \glspl{dop-op} necessary to support
a new application,
because they usually only serve to correct the oddities of the
underlying \glspl{dop-gadget}.
Wireshark in particular has to deal with this,
which explains the high count of application-specific \glspl{dop-op}.
The exceptionally high line count for these \glspl{dop-op} for the Mini-Server
is caused by a single outlier: The \lstinline{store-add}-to-\lstinline{store}-gadget
requires 72 lines.

Lastly, there are reusable \glspl{dop-op}.
They use only properly working \glspl{dop-op} and are
thus fairly easy to implement, despite using more lines.
As Table~\ref{table-nof-gadgets} shows,
one needs roughly a dozen of these to execute all the \glspl{dop-program}
from our experiments on a particular target application.

More importantly, Table~\ref{table-nof-gadgets} also shows that,
if one keeps the distinction between these three categories in mind,
e.\,g., with a naming convention to rule out accidental mix-ups,
one can reuse many \glspl{dop-op}.
For example, four of the five target applications share the recipe for
\lstinline{mul} and \lstinline{sub}.
All in all, about 75\,\% of the \glspl{dop-op} without application-specific
oddities actually are reused.
Also note the high count for the Mini-Server and ProFTPd.
Both require the branch-free mode,
which in turn requires gadgets that only have an effect,
if the \textit{here}-bit of a basic block is set.
Not a single \gls{dop-op} from this category had to be implemented additionally
to support ProFTPd.

In summary,
to support four algorithms,
we needed about 15-20 \glspl{recipe} per application, each having roughly 10 lines.
While they may require domain knowledge to circumvent the app's peculiarities,
writing them is (roughly) as simple as writing assembly.
We found that about half of the \glspl{recipe} can be reused from other applications,
and that they furthermore share certain patterns,
which makes writing new \glspl{recipe} easier.
Over time, we expect that more recipes can be reused and that therefore
less recipes need to be implemented from scratch.
Finally, note that an attacker is likely to only implement
a very limited number of \glspl{dop-gadget} %
for one specific payload.

\input{tables/nof-gadgets}

%% file: tables/table-non-interactive.tex
\newcommand{\STAB}[1]{\begin{tabular}{@{}c@{}}#1\end{tabular}}

\begin{table*}
  \centering
  \caption{Evaluation of \glspl{dop-program} compiled for three non-interactive applications: Interpreter, GStreamer, and Wireshark.\\
1: Optimizations: Loop unrolling; using variables instead of array.
}
\rotatebox[origin=c]{90}{\textbf{\hspace{0.2cm}GStreamer\hspace{0.8cm}Wireshark\hspace{0.6cm}Interpreter\hspace{0.6cm}}}
  \begin{tabular}{l|r|r|r|r|r}
  & Levenshtein & SSL-Deref       & Relocator      & JIT-ROP    & $\text{JIT-ROP}^1$ \\
\glspl{dop-gadget} & 183         & 17              & 23             & 99         & 56              \\
\hline
Compile Time          & 23.14s      & 1.53s           & 1.61s          & 12.11s     & 1.85s           \\
... per \gls{dop-gadget}& 0.12s       & 0.09s           & 0.07s          & 0.12s      & 0.03s           \\
\hline
Execution Time        & 0.29s       & 0.19s           & 0.20s          & 0.51s      & 0.49s           \\
... per \gls{dop-gadget}& 0.50$\mu$s  & 1898.89$\mu$s   & 2115.79$\mu$s  & 0.14$\mu$s & 0.35$\mu$s      \\
\hline
Executed \glspl{dop-gadget}& 5855        & 99              & 95             & 35982      & 14180           \\
\hline
\hline
\glspl{dop-gadget} & 4407        & 1930            & 1419        & 2503          & 2689            \\
\hline                                                                                                
Compile Time          & 179.75s     & 108.96s         & 91.62s      & 133.37s       & 107.89s         \\
... per \gls{dop-gadget}& 0.04s       & 0.06s           & 0.06s       & 0.05s         & 0.04s           \\
\hline                                                                                                
Execution Time        & 8.43s       & 0.75s           & 0.70s       & 77.85s        & 13.39s          \\
... per \gls{dop-gadget}& 2.81$\mu$s  & 3.59$\mu$s      & 3.84$\mu$s  & 2.91$\mu$s    & 2.85$\mu$s      \\
\hline                                                                                                
Executed \glspl{dop-gadget}& 2,994,468   & 210,989         & 195,758     & 26,733,658    & 4,691,568       \\
\hline
\hline
\glspl{dop-gadget} & 2284        & 265            & 317        & 740          & 1079            \\
\hline                                                                                                
Compile Time          & 467.89s     & 59.72s         & 102.47s      & 227.07s       & 334.75s         \\
... per \gls{dop-gadget}& 0.20s       & 0.22s           & 0.36s       & 0.31s         & 0.31s           \\
\hline                                                                                                
Execution Time        & 16.33s       & 1.45s           & 1.59s       & 228.50s        & 149.58s          \\
... per \gls{dop-gadget}& 84.75$\mu$s  & 728.91$\mu$s      & 447.12$\mu$s  & 86.52$\mu$s    & 81.03$\mu$s      \\
\hline                                                                                                
Executed \glspl{dop-gadget}& 192,676   & 1,992         & 3,565     & 2,640,980    & 1,845,845       \\
\hline
  \end{tabular}
  \label{table-quantitative-non-interactive}
\end{table*}

%% file: tables/table-interactive.tex
\begin{table*}
  \centering
  \caption{Evaluation of \glspl{dop-program} compiled for two interactive applications: Mini-Server and ProFTPd.\\
1: Optimization: Loop-Check against constant.\\
2: Optimization: Loop unrolling; no branch-free transformation, since it is only one Basic Block.\\
3: Optimization: Using a single 16-bit-compare instead of two 8-bit compares.
\hfill
*: Estimated.
}
\rotatebox[origin=c]{90}{\textbf{ProFTPd\hspace{0.8cm}Mini-Server\hspace{0.4cm}}\vspace{3mm}}
  \begin{tabular}{l|r|r|r|r|r|r|r|r}
                      & Levenshtein & $\text{Levenstein}^1$ & SSL-Deref & $\text{SSL-Deref}^2$ & Relocator & $\text{Relocator}^2$ & $\text{JIT-ROP}^3$ & $\text{JIT-ROP}^1$ \\
\glspl{dop-gadget}    & 5466        & 4789                  & 762       & 141                  & 1112      & 37                   & 594                & 316      \\
\hline
Compile Time          & 807.37s     & 686.47s               & 93.91s    & 17.68s               & 117.43s   & 10.86s               & 80.54s             & 50.40s  \\
... per \gls{dop-gadget} & 0.14s       & 0.14s                 & 0.12s     & 0.12s                & 0.11s     & 0.29s                & 0.14s              & 0.16s   \\
\hline
Execution Time        & 5360.64s    & 420.36s               & 14.71s    & 0.62s                & 61.48s    & 0.25s                & *22,956.00s        & 190.67s \\
... per \gls{dop-gadget} & 1.28ms      & 1.28ms                & 1.39ms    & 4.46ms               & 1.28ms    & 6.83ms               & *1.28ms            & 1.28ms  \\
\hline
Executed \glspl{dop-gadget} & 4,178,599   & 327,671               & 11,269    & 141                  & 47,921    & 37                   & 17,894,894         & 148,625  \\
\hline
\hline
\glspl{dop-gadget}    & 2946        & 2813                  & 354       & 126                  & 607       & 72                   & 501                & 195      \\
\hline
Compile Time          & 471.37s     & 447.27s               & 59.83s    & 23.56s               & 98.94s    & 21.74s               & 84.17s             & 33.15s  \\
... per \gls{dop-gadget} & 0.16s       & 0.16s                 & 0.17s     & 0.19s                & 0.16s     & 0.30s                & 0.17s              & 0.17s   \\
\hline
Execution Time        & 3245.78s    & 1641.69s              & 6.93s     & 0.59s                & 49.21s    & 0.56s                & *20,008.00s        & 123.95s \\
... per \gls{dop-gadget}& 1.31ms      & 1.31ms                & 1.32ms    & 4.69ms               & 1.31ms    & 7.18ms               & *1.31ms            & 1.31ms  \\
\hline
Executed \glspl{dop-gadget} & 2,472,032   & 1,250,336             & 5,243     & 126                  & 37,452    & 72                   & 15,238,408         & 96,406   \\
\hline
  \end{tabular}
  \label{table-quantitative-interactive}
\end{table*}

%% file: tables/nof-gadgets.tex
\begin{table}
  \centering
  \caption{Number of implemented \glspl{dop-op}.\\
Different operand sizes were ignored for this table, as recipes for conversion are all reusable,
and usually just use temporary variables to zero-extend or save parts of the operands.}
  \begin{tabular}{l|c|c|c|c|c}
                              & \rotatebox{63}{Interpreter} & \rotatebox{63}{Wireshark} & \rotatebox{63}{GStreamer} & \rotatebox{63}{Mini-Server} & \rotatebox{63}{ProFTPd} \\
\hline
As \Glspl{gadget-definition}   &  6                          &  5                        &  1                        &  5                          &  5                      \\
\hline
$\emptyset$ lines of code     &  3.8                        &  14                       &  9                        &  5                          &  11.2                   \\
\hline
\hline
App-specific                  &  3                          &  8                        &  6                        &  6                          &  5                      \\
\hline
$\emptyset$ lines of code     &  2.6                        &  6.6                      &  4.2                      &  18.6                       &  5.6                    \\
\hline
\hline
Reusable                      &  5                          &  11                       &  11                       &  16                         &  11                     \\
\hline
$\emptyset$ lines of code     &  6.2                        &  10.5                     &  8.3                      &  4.7                        &  4.6                    \\
\hline
Shared                        &  5                          &  5                        &  8                        &  13                         &  11                     \\
\hline
  \end{tabular}
  \label{table-nof-gadgets}
\end{table}

%% file: sections/discussion.tex
\section{Discussion}
In this section, we discuss the security implications of the possibility
to write complex \gls{dop} programs, limitations of our approach,
and possible future work.

\subsection{DOP Expressiveness}
Our experiments have clearly shown that more powerful \gls{dop} instances
lead to more efficient \glspl{dop-program}.
Still, it is also clear that remarkably little is necessary for \gls{dop}.
E.\,g., our branch-free transformation not only shows that the interactive mode
is just as expressive as the non-interactive mode,
but also that neither a virtual program counter nor conditional \glspl{dop-gadget}
are necessary.

Furthermore, we have shown in the Interpreter example,
that the data-movement \glspl{dop-gadget} can express one another,
i.\,e., that having either
\lstinline{mov} or \lstinline{load} or \lstinline{store} can be sufficient.
The Wireshark example again lowers the bar for Turing-complete
\gls{dop} by showing that one can express arithmetic gadgets through data-movement,
while the Mini-Server example shows that data-movement gadgets can express arithmetics.
Lastly, none of our exemplary real-world target applications has a native logical gadget.

\subsection{System Interaction}
I/O is certainly useful,
but we argue that arbitrary computations in another program's memory are useful on its own,
e.\,g., to change the program's state or to compute values for further attack steps,
like the \glspl{dop-script} in our evaluation.
Naturally, DOP can only interact with the system,
if the necessary system calls happen to be reachable via data flows.
In this case, a \gls{recipe} for a system call \gls{dop-gadget} could be crafted.
Alternatively, one would have to alter the control flow,
i.\,e., use a \gls{recipe} to wrap redirection, system call and return.
While this could certainly be facilitated by \gls{dop},
one would cross the border of pure \gls{dop}.
Similar to traditional code-reuse, an attacker is much more likely
to use \gls{dop} to bootstrap an additional attack-phase
to accomplish the chosen task in an easier way.

\subsection{Security Implications}
Our exemplary \glspl{dop-script}
demonstrate how \gls{dop} and \gls{byose} can aid in
leaking sensitive memory,
in relocating a \gls{rop-chain} to bypass coarse-grained code-layout randomization,
and in finding \glspl{rop-gadget} on-the-fly to bypass fine-grained code-layout randomization.
However, there are other attacks against more powerful
defense primitives, which also require a scripting environment.
Thus, applications lacking a scripting environment are immune to those
attacks, unless, of course, \gls{dop} is possible.
E.\,g., one can
bypass coarse-grained \gls{cfi}~\cite{out-of-control}
by probing the memory for \glspl{rop-gadget} still usable even with the
\gls{cfi} policy in place.
Evans et al.~\cite{missing-the-pointer} bypassed the
64-bit version of \gls{cpi}, which protects sensitive pointers,
by efficiently locating the safe-region it uses to hide its metadata.
Furthermore, variants of \gls{xom} are vulnerable to runtime attacks deducting
not-yet-protected bytes~\cite{bgdx}.

We believe the recent progression of \gls{doa} and \gls{dop} indicate that the research community should
consider protecting data-flow and data structures when designing new systems defenses.

\subsection{Limitations}
Our research prototype has several limitations, which we want to discuss here
together with their implications.

\subsubsection{Data-Pointers}
Our attacks may require a few data pointers or offsets as input,
which may have to be leaked or guessed first.
It can be argued that there are situations, where this is not excluded
by the presence of ASLR (e.\,g., static modules, info leaks, process forks
using the same memory layout, incompatible defenses, etc.),
but it means that our exploits technically may not
circumvent traditional ASLR.
However, regarding code-reuse attacks,
defenses like fine-grained code randomization~\cite{binary-stirring}
would be a stronger drop-in replacement for ASLR,
but since they do not need to modify the data layout
they are likely not to require such hard-to-guess data pointers.
Furthermore,
analogously to attacks targeting code-pointers~\cite{pirop},
the GStreamer example shows that there are applications,
in which relative offsets instead of absolute addresses
suffice for \gls{dop}-attacks.

\subsubsection{End-to-End Exploitation}
While we deem this work to be a substantial improvement in automating
\gls{dop}, it does not result in the fully automated construction of end-to-end \gls{dop} exploits.
\outline{app-summary to auto-generated}
Firstly, the definitions of an application's \glspl{dop-gadget}
are not generated automatically.
While we, conceptually, use the output of previous work~\cite{dop},
the format and details of said definition require manual effort.
\outline{driver not auto-generated}
Secondly, embedding the data structure in the application's input
and interfacing the application is left to the attacker,
although our tools automatically generate the invocations to the interface.

\subsubsection{Optimizations}
Our prototype \tool{} uses little traditional compiler
optimization, like constant folding or common subexpressions reuse.
However,
in exploitation,
a slightly better or smaller program may not merely be faster,
but increase the chance of success substantially.

\subsection{Future Work}
Naturally, many technical limitations mentioned above can be seen
as future work.
However, we also want to discuss what we feel to be conceptual gaps in our
understanding of \gls{dop}.

\subsubsection{Confined and arbitrary code-execution}
Further examination of the gap between \gls{dop}
and arbitrary code-execution may be worthwhile.
E.\,g., \gls{dop} may enable Turing-complete computation,
but its results can not always be saved in a suitable format
to interact with the outside of the \gls{dop} bubble.

\subsubsection{Multi-Stage Bootstrapping}
Our prototype does not reason about incomplete or constrained \glspl{dop-gadget},
but instead relies on the attacker to create definitions or recipes for
sufficient \glspl{dop-gadget}.
Similarly, the possibilities of executing multiple \glspl{dop-gadget} in one step,
like the \lstinline{mov-add} to implement the \lstinline{store} for Wireshark
or the double-\lstinline{mov} to implement the \lstinline{store/load} for ProFTPd,
are yet to be fully explored.
An adaptive interactive mode, which uses info leaks to adapt the sequence or
content of following \glspl{dop-gadget} would also pose an interesting opportunity.
If the search for \glspl{dop-gadget} would support such bootstrapping,
we suspect that
many more applications would allow \gls{byose}-attacks.

\subsubsection{Defenses}
Concrete \gls{dop} attacks can be prevented
in many ways: Fixing underlying vulnerabilities,
data layout randomization, data obfuscation,
or constrained read-/write-targets~\cite{wit,hardscope}.
Since \gls{dop} is arguably even more complex than \gls{rop},
there are likely even more subtle ways to hamper attackers.
However, it remains a challenge to create efficient, easy-to-apply defenses,
which
constricts the very concept of \gls{dop}
and %
\glspl{doa}.

%% file: sections/related-work.tex
\section{Related Work}

This section %
provides an overview over
orthogonal attacks threatening modern defense mechanisms,
the state of the art for data-only attacks,
and defenses trying to prevent data-only attacks.

\subsection{Orthogonal Attacks}
\Gls{coop}~\cite{coop} achieves ROP-like
capabilities using only full-function gadgets comprised of the methods of
suitably crafted objects,
without violating the non-C++-specific \Gls{cfg}.
Crash Resistance~\cite{crash-resistance} allows to probe memory,
even though a careless probe should crash the application.
In combination with a full-function reuse technique,
this can thwart code randomization and schemes for information hiding.
Blind ROP~\cite{blind-rop} uses a side-channel to locate gadgets in an
otherwise unknown binary with a stack-buffer vulnerability.
This requires that the attacked server-application is restarted after a crash,
without being rerandomized.
\textit{Control-Flow Bending}~\cite{control-flow-bending} is a code-reuse
attack that exploits the coarse-grained nature of some CFI implementations.
They chain pairs of calls to standard library functions such as \texttt{memcpy/printf}
and control their arguments to achieve Turing-complete computation on the data plane.

Conti et al.~\cite{losing-control} show that current
CFI- %
and Shadow Stack-implementations are vulnerable
to attackers controlling stack-values though a heap-based vulnerability.
G\"{o}ktas et al.~\cite{out-of-control} show that coarse-grained \gls{cfi} does
not prevent the execution of call-site gadgets
and entry-point gadgets. %
Control Jujutsu~\cite{control-jujutsu}
can circumvent even fine-grained CFI with
a shadow stack,
because modern applications often
use coding practices,
which exceed the limits of current %
pointer analysis.
Snow et al.\cite{zombie-gadgets} presented multiple attacks highlighting
implementation pitfalls of \gls{dcr}-based defenses,
whereas BGDX~\cite{bgdx} shows a more general attack deducing the location
of not-yet-protected gadgets.

\textit{Q}~\cite{Q-exploit-hardening} is conceptually similar to this work
in that they provide a language for exploit programming, too.
The key difference is that
a) their language is more low-level than ~\ourlanguage{},
b) their languages is based on %
\glspl{rop-gadget}, which
are much more regular across applications and
easier to reuse.

The authors of \textit{Microgadgets}~\cite{microgadgets} define classes of \glspl{rop-gadget},
and, trying to use those members with the fewest bytes,
implement higher-level operations.
E.\,g., using multiple \lstinline{xor}-gadgets to implement a \lstinline{store},
which is similar in concept to our recipes,
but without supporting the automatic selection or recombination we achieve.

\subsection{Data-Only Attacks}
Chen et al.~\cite{data-only} have shown that %
attacking non-control-data can have consequences just as severe as code-reuse attacks.
\textit{Memory Cartography}~\cite{new-doa} is an automatic way to create
a net of memory references to navigate reliably through data structures,
and can be
used it to create \gls{doe}, which are robust to
ASLR.
Hu et al. presented a technique called \textit{dataflow stitching}~\cite{automatic-data-oriented},
which analyses potentially corruptible data-flow in an application,
especially with regard on how to chain individual data-flows.
However, their attacks are targeted to enable specific attacks targeting
the peculiarities of a particular application, much in the spirit of Chen et al..
Based on these stitching techniques, however, they recently developed what is
now called \gls{dop}~\cite{dop}.
The authors not only show that \gls{dop} often allows Turing-complete computation,
but also that \glspl{dop-gadget} are frequently available.
However, they focus mainly on gadget search, and leave both the DOP instance
setup and the payload preparation to manual work.

Block-Oriented Programming~\cite{block-oriented} accepts a
script in a high-level language and a read/write-primitive as input.
In a nutshell, 
it then tries to solve the constraints a given program has,
to ultimately synthesize the script using data-oriented techniques.
It does so with a high degree of automation,
but the authors also show that their approach of program synthesis via
constraint solving is NP-hard.
Thus, the synthesized programs are rather simple,
essentially only making sure that some data ends up in a sensitive sink.
We would argue that these programs still belong to the step of DOP instance setup.

In contrast, this paper is mostly not concerned with DOP \textit{gadget search},
or the \textit{DOP instance setup},
but instead focuses on the
\textit{payload preparation}, which is
automating %
the process of creating the data structures
that drive an application to execute the many instructions of a \gls{dop-program}.

\subsection{Defenses against Data-Oriented Attacks}
Current countermeasures against data-flow attacks seem to mirror those against
control-flow attacks.
Lin et al.~\cite{randomizing-data-structure} and \textit{SALADS}~\cite{salads}
randomize the layout of data structures.
Both \textit{data-randomization}~\cite{data-randomization}
and \textit{Data Space Randomization}~\cite{data-space-randomization}
masks data in-between uses by XOR-ing them with random values.
The authors of \textit{HARD}~\cite{hard} take this concept to the hardware-level,
with a customized RISC-V architecture.
These five techniques are based on introducing randomness,
but may not protect all data and furthermore require source code,
which makes them unsuitable for protecting \gls{cots} binaries.
\textit{ValueGuard}~\cite{valueguard} inserts canary values in-between buffers and other data
to
detect a buffer overflow. %
Bhatkar et al.~\cite{comprehensive-memory-error},
shuffle functions, randomize heap object-locations and location of the stack,
and use separate stacks for stack buffers.
Both approaches significantly hamper \glspl{doa}, but may not protect against
such attacks in all cases (especially if they are not based on buffer
overflows).

Analogous to \Gls{cfi}, \Gls{dfi}~\cite{data-flow-integrity} uses a statically
determined data-flow graph to restrict the data-flow at runtime.
\Gls{wit}~\cite{wit} analyses a program
to color instructions and memory objects.
It then ensures at runtime that instructions only write to objects with a matching color.
\textit{CUP}~\cite{cup} combines
memory-maps with fat-pointers to ensure
spatial and (probabilistic) temporal memory safety, even for stack-variables.
\textit{HardScope}~\cite{hardscope} carries the concept of a variable's visibility-scope
from the source code to the runtime.
These techniques incur high overhead,
even despite the fact that \textit{HardScope} is implemented with hardware
support on an open-source microcontroller.

Exploits are naturally brittle, so we would expect most defenses to
thwart most unmodified \glspl{dop-exploit}.
In comparison to code-flow transfers,
data is very dynamic and touched virtually everywhere in a program,
which makes it hard to reason about and inefficient to check its integrity.
Thus, we expect that it will be even harder to create efficient defenses,
which prevent \glspl{doa} in principle rather than by chance.
E.\,g., the authors of \textit{HardScope} acknowledge that
preventing the advanced exploit against GStreamer~\cite{gstreamer-exploit}
that we also use in our experiments is more challenging,
because it corrupts a heap object it could use legitimately as well.
Similar to CFI bypasses, some \glspl{doa} may fall through the cracks,
e.\,g., due to unprotected modules, false metadata, or coarse class-granularity.

%% file: sections/conclusion.tex
\section{Conclusion}

In the last decade code-reuse defenses gained a lot of attention in academia and industry,
which led to their wide-spread adoption.
However,
the mitigations of the orthogonal \gls{doa} and \gls{dop} attacks are still in the early stages of development.

In this paper we demonstrate that even constrained \gls{dop} instances can be escalated in expressiveness to execute complex \glspl{dop-program}. We show that \glspl{dop-program} aid in bypassing code-reuse defenses to launch advanced code-reuse attacks such as JIT-ROP. This BYOSE feature of \gls{dop} transfers a whole class of just-in-time attacks to targets without a built-in scripting engine. Our high-level~\ourlanguage{} language can be used to implement portable \gls{dop} exploits and our \gls{dop} compiler~\tool{} automatically constructs the required low-level data structures to run them in target applications.
We hope that our results raise the awareness for \gls{doa} and \gls{dop} in the research community and aid in the development of systems defenses that consider the mitigation of these attacks.

%% file: sections/acknowledgements.tex
\noindent
This work was supported by
the European Research Council (ERC) under the European Union's
Horizon 2020 research and innovation programme (ERC Starting
Grant No. 640110 BASTION). In addition, this work was 
supported by the German Federal Ministry of Education and Research (BMBF Grant 16KIS0592K HWSec).